\documentclass[10pt,journal,twoside,onecolumn]{IEEEtran}

\usepackage{epsfig,graphics}
\usepackage{amsmath,amssymb,dsfont,multirow}
\usepackage{subfigure,url}%,ulem}
  \usepackage[below]{placeins}
  \usepackage{cite}
  \usepackage[ruled,vlined]{algorithm2e}

  \usepackage[utf8]{inputenc}
\usepackage[T1]{fontenc}
 \usepackage[table,xcdraw]{xcolor}

\newcommand\blfootnote[1]{%
  \begingroup
  \renewcommand\thefootnote{}\footnote{#1}%
  \addtocounter{footnote}{-1}%
  \endgroup
}
 \usepackage{empheq} 

  \usepackage{relsize}
  \usepackage{comment}
\usepackage[figuresright]{rotating}

\usepackage{color}
\definecolor{agreen}{rgb}{0,0.5,0}

\definecolor{bkblue}{HTML}{DAE8FC} 
\definecolor{bdblue}{HTML}{184686}

\newcommand{\cp}[1]{\ifmmode {\mathcal{#1}}\else ${\mathcal{#1}}$\fi}
\newcommand{\balpha}{{\boldsymbol{a}}}

\newcommand{\beps}{\boldsymbol{\epsilon}}

\newcommand{\ba}{\boldsymbol{a}}
\newcommand{\bb}{\boldsymbol{b}}

\newcommand{\bh}{\boldsymbol{h}}

\newcommand{\bm}{\boldsymbol{m}}
\newcommand{\bw}{\boldsymbol{w}}
\newcommand{\bx}{\boldsymbol{x}}
\newcommand{\by}{\boldsymbol{y}}
\newcommand{\bz}{\boldsymbol{z}}
\newcommand{\bv}{\boldsymbol{v}}
\newcommand{\bW}{\boldsymbol{W}}

\newcommand{\bB}{\boldsymbol{B}}

\newcommand{\bA}{\boldsymbol{A}}

\newcommand{\bU}{\boldsymbol{U}}
\newcommand{\bV}{\boldsymbol{V}}
\newcommand{\bM}{\boldsymbol{M}}
\newcommand{\bY}{\boldsymbol{Y}}
\newcommand{\bZ}{\boldsymbol{Z}}

\newcommand{\bTheta}{\boldsymbol{\Theta}}

\newcommand{\bI}{\boldsymbol{I}}

\newcommand{\boxedeq}[2]{\begin{empheq}[box={\fboxsep=6pt\fbox}]{align}\label{#1}#2\end{empheq}}

\newcounter{ExpNo}
\stepcounter{ExpNo}

\begin{document}

\title{Integration of Physics-Based and Data-Driven Models for Hyperspectral Image Unmixing}

\author{Jie Chen$^{\star}$, \IEEEmembership{Senior Member, IEEE}, Min Zhao$^{\star}$, \IEEEmembership{Student Member, IEEE}, Xiuheng Wang$^{\dag}$, \IEEEmembership{Student Member, IEEE} \\
C{\'e}dric Richard$^\dag$, \IEEEmembership{Senior Member, IEEE}, Susanto Rahardja$^{\star{\ddagger}}$, \IEEEmembership{Fellow, IEEE}
 \\ \vspace{0.25cm}
 \small{\linespread{0.1} $^\star$ %Center of Intelligent Acoustics and Immersive Communications \\
School of Marine Science and Technology, Northwestern Polytechnical University, 710072, Xi'an, China \\
dr.jie.chen@ieee.org, minzhao@mail.nwpu.edu.cn, susantorahardja@ieee.org} \vspace{0.2cm}\\
\small{\linespread{0.1} $^\dag$ Laboratoire Lagrange,  Universit{\'e} C\^ote d'Azur,  06000, Nice, France \\
xiuheng.wang@oca.eu, cedric.richard@unice.fr}\\
\small{\linespread{0.1} $^\ddagger$ {Singapore Institute of Technology, 138683, Singapore}}
}

\maketitle

\vspace{-1.4cm}

%=================================
% Motivation and significance of the topic
%=================================
%\newpage
\begin{abstract}
    Spectral unmixing is central when analyzing hyperspectral data. To accomplish this task, physics-based methods have become popular because, with their explicit mixing models, they can provide a clear interpretation. Nevertheless, due to their limited modeling capabilities, especially when analyzing real scenes with unknown complex physical properties, these methods may not be accurate. 
    %Data-driven methods have developed rapidly in recent years, in particular deep learning methods as they possess superior capability in modeling complex and nonlinear systems. 
    {On the other hand, data-driven methods using deep learning in particular have developed rapidly in recent years thanks to their superior capability in modeling complex nonlinear systems.}
    Simply transferring these methods as black boxes to perform unmixing may lead to low interpretability and poor generalization ability. To bring together the best of two worlds, recent research efforts have focused on combining the advantages of both physics-based models and data-driven methods. In this article, we present an overview of recent advances on this topic from various perspectives, including deep neural networks (DNN) design, prior capturing and loss selection. We summarise these methods within a common optimization {framework and discuss ways for enhancing the understanding of these methods.}
    %framework. In addition, discussions are conducted for enhancing the understanding of these methods and suggest prospective improvements. 
    The related source codes are made publicly available\footnote{ {http://github.com/xiuheng-wang/awesome-hyperspectral-image-unmixing}}.

\end{abstract}

\section{{Introduction}}

%{Conventional machine vision applications widely use Red-Green-Blue (RGB) cameras. These cameras are suitable for identifying objects with shapes and colors that are consistent with human eyes and visual perception. As a significant advance in the imaging techniques, hyperspectral cameras measure objects or scenes by recording hundreds of narrow bands across a wide-range of contiguous spectrum which can cover non-visible range of the light.} Benefiting from their rich spectral information coupled with regular spatial information, hyperspectral images allow identifying materials based on their unique spectral signatures beyond visible characteristics. {The initial applications of hyperspectral imaging focused on remote sensing of the earth surface using satellite or air supported cameras for mineral exploration, vegetation monitoring and land analysis. The subsequent technical developments of hyperspectral cameras and the ongoing cost decrease  allow us now to investigate their use in industrial, medical and civil applications. Specific emerging applications include food safety inspection, medical diagnostics, industrial sorting, drug analysis, biometric forensics, archaeology, etc.}

Conventional machine vision applications widely use Red-Green-Blue (RGB) cameras. These cameras are suitable for identifying objects with shapes and colors consistent with the human visual system. As a significant improvement in imaging techniques, hyperspectral cameras measure objects or scenes by recording hundreds of narrow bands across a wide-range spectrum that can cover the non-visible range of light. Hyperspectral images then enable the identification of materials based on their unique spectral signatures, beyond their visible {characteristics,} coupled with spatial information~\cite{ghamisi2017advances}. Initial hyperspectral imaging applications {started with} remote sensing of the Earth's surface using satellite or airborne cameras for mineral exploration, vegetation monitoring, and land analysis. The subsequent technical advancements and ongoing cost reduction of hyperspectral cameras {have made their deployment practical and possible in other} emerging applications such as food safety inspection, medical diagnostics, industrial sorting, drug analysis, biometric forensics, archaeology, etc.

\begin{figure*}[!t]
  \centering
  % Requires \usepackage{graphicx}
  \includegraphics[width=16cm]{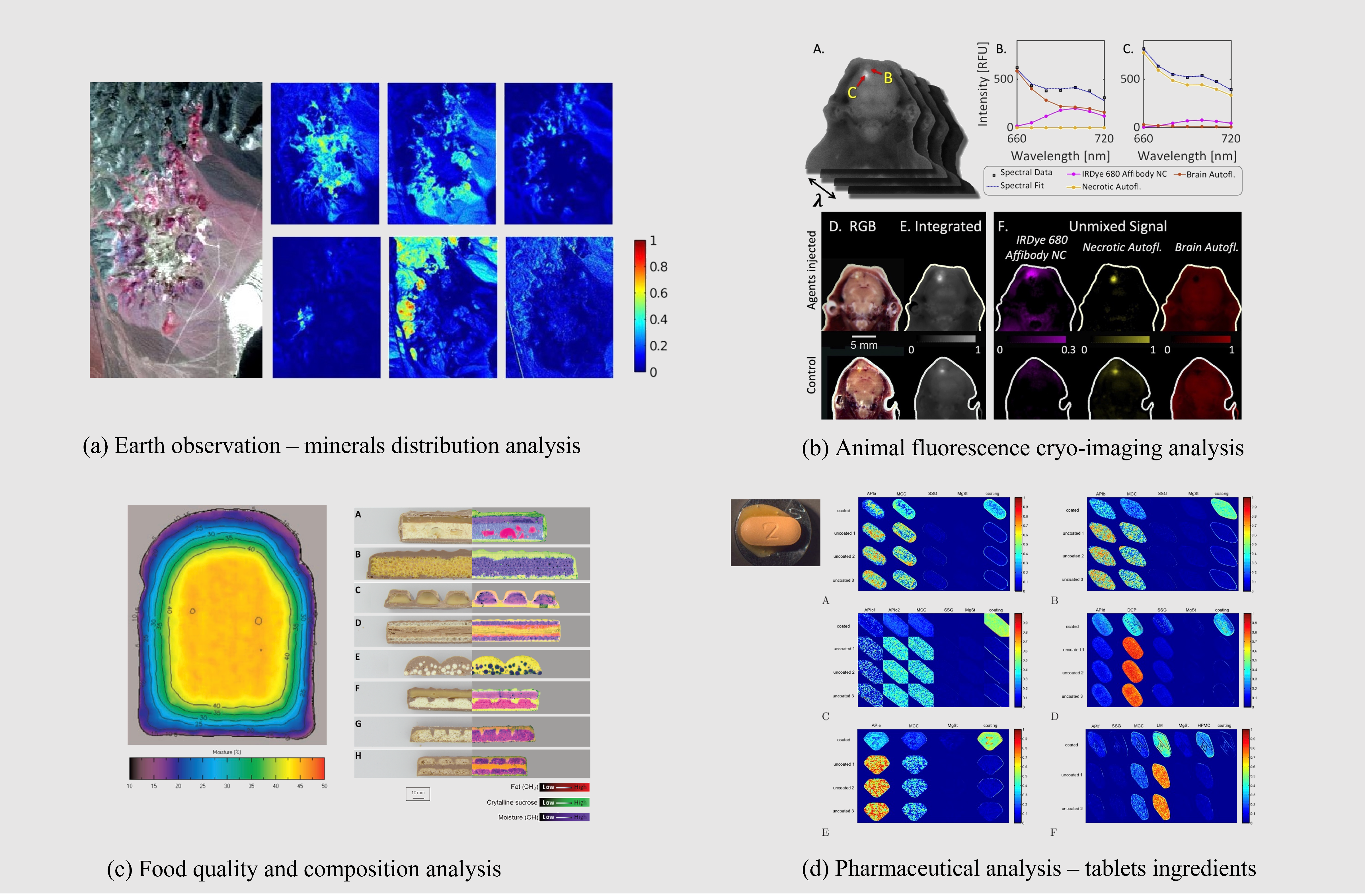}
    \vspace{-3mm}
  \caption{{Applications of hyperspectral data unmixing. (a) Analysis of mineral distribution of remotely sensed data (CUPRITE)\cite{chen2013nonlinear_sp}; (b) Spectral unmixing for improving whole-body fluorescence cryo-imaging$^1$; (c) Food analysis: moisture distribution in a fresh slice of bread and composition analysis of commercial chocolate bars$^2$; (d) Pharmaceutical analysis: tablets ingredients distribution estimation$^3$. }}
  \label{fig_unmixingapplication}
  \vspace{-5mm}
\end{figure*}

The spectral content of individual pixels in hyperspectral images is typically a mixture of the spectra of multiple materials~\cite{dobigeon2013nonlinear,borsoi2021spectral}. This phenomenon called spectral mixture is attributed to multiple factors, mainly including the low spatial resolution of hyperspectral imaging devices, the diversity and intimate interactions of materials in the imaged scenes, and multiple photon reflections from layered objects. Separating spectra of individual pixels into a set of spectral signatures (called endmembers), and determining the abundance fraction of each endmember, is an essential task in quantitative hyperspectral subpixel analysis. This process, denoted as \emph{spectral unmixing} or mixed pixel {decomposition, is currently used in many applications such as} conventional mineral distribution analysis in remote sensing, biodistribution analysis in fluorescence images, food quality and composition control, and pharmaceutical inspection, as illustrated in Fig.~\ref{fig_unmixingapplication}.

\blfootnote{$^{1}$\url{https://opg.optica.org/boe/fulltext.cfm?uri=boe-12-1-395&id=444823}}
\blfootnote{$^{2}$\url{https://www.specim.fi/better-control-of-food-quality-and-composition-with-hyperspectral-imaging}}
\blfootnote{$^{3}$\url{http://www.lx.it.pt/~bioucas/files/PhD_Thesis_MBLopes.pdf}}

\begin{table}[!t]
\caption{Main symbols used in the paper.}
\label{tab_notation}
\centering
\begin{tabular}{
>{\columncolor[HTML]{DAE8FC}}l 
>{\columncolor[HTML]{DAE8FC}}l| 
>{\columncolor[HTML]{DAE8FC}}l 
>{\columncolor[HTML]{DAE8FC}}l }
\hline\hline
$x, X$  & scalar & $\boldsymbol{y}$  & a general observed pixel \\\hline
$\boldsymbol{x}$  & column vector & $\boldsymbol{y}_i$ & $i$-th pixel of the image, {with $1\le i \le N$}\\\hline
$\boldsymbol{X}$ & matrix  & $\boldsymbol{m}_i=[m_{i,1}, \cdots, m_{i,L}]^{\top}$ &  $i$-th endmember, {with $1\le i \le R$}\\ \hline
$\mathbb{X}$ & 3D data cube  & $\boldsymbol{M}= [\boldsymbol{m}_1,\cdots, \boldsymbol{m}_{R}]$ &  endmember matrix with $R$ pure spectra \\ \hline
$N$  & {number of pixels} & $\balpha=[{a}_1, \cdots, {a}_R]^{\top}$ & a general abundance vector \\ \hline
$L$  & {number of spectral bands} & $\balpha_i=[{a}_{i,1}, \cdots, {a}_{i,R}]^{\top}$ & abundance vector of the $i$-th pixel \\ \hline
$R$  & {number of pure spectra (endmembers)} & $\boldsymbol{Y}=[\boldsymbol{y}_{1}, \cdots, \boldsymbol{y}_{N}]$ & matrix by stacking $\mathbb{Y}$ with {$N$ pixels} \\ \hline
$\mathbb{Y}$ & image with {$N$} pixels and $L$ spectral bands   & $\boldsymbol{A}= [\balpha_1,\cdots, \balpha_N]$ & abundance matrix of {$N$ pixels}
  \\\hline\hline
\end{tabular}
\end{table}
Diverse solutions have been proposed for the spectral unmixing problem. They can be divided primarily into two categories: physics-based and data-driven methods. Physics-based methods exploit physical processes of light scattering and interaction mechanisms among multiple materials in a scene.
%interacting with multiple scene materials. 
The observed spectra are explicitly related to the endmembers and abundances, albeit with strong assumptions regarding photon-interacting processes. In practice, it is difficult to accurately model real scenes, and sophisticated models {with complex mathematics are usually difficult to solve.}
%contain mathematics that are difficult to manipulate. 
In contrast, data-driven unmixing methods can overcome some limitations {of the aforementioned} physics-based methods by directly deriving mixture models from the observed hyperspectral data. These techniques are effective at uncovering the underlying relationships in complicated scenes with abundant data, however, explicit and useful knowledge of the physical mixing process is frequently overlooked in naive data-driven approaches. In order to reap the benefits of both physics-based and data-driven models, it is essential to properly integrate these techniques in order to achieve superior unmixing performance with clear interpretation.

The main purpose of this article is to provide an overview of hyperspectral unmixing models and techniques with special attention focused on the integration of physics-based models and data-driven methods. These methods are analyzed from multiple perspectives, including model structure design, prior capturing, and loss function design. In addition to summarizing the materials dispersed throughout the literature, we establish a mathematical framework that can characterize and relate these methods. {This article provides a set of references that are sufficient to cover the contributions mentioned and the methods introduced here should also serve as good examples for understanding the topics.}
%By no means does this article include a complete list of references for all contributions, as the field is growing quickly, but the methods introduced here should serve as good examples for understanding the topic. 
The main symbols used in this article are provided in Table~\ref{tab_notation}.

\section{A general problem formulation of the unmixing problem}

This section establishes a general modeling and optimization framework for spectral unmixing. The classical and recent physics-based and data-driven unmixing methods described {subsequently} are compatible with this framework.

{Generally, hypersepctral unmixing consists of three essential steps namely,}
%Hypersepctral unmixing consists of three essential steps, namely,
estimation of the number of endmembers, endmember extraction and abundance estimation. Estimation of the number of endmembers can be performed with signal dimension estimation techniques or circumvented with sparse regression. Endmember estimation and abundance estimation can be processed sequentially or simultaneously. In this process, endmembers and abundances are frequently constrained to ensure their physical interpretability. {Endmembers represent the pure spectra existing in the image, implying that the endmember nonnegativity constraint (ENC) should be considered. Abundances represent the contribution in percentage of each pure component, the abundance nonnegativity constraint (ANC) and the abundance sum-to-one constraint (ASC) are often considered.} The feasible region of endmembers and abundances is denoted by $\Omega_M$ and $\Omega_{a}$, with:
 \begin{align}
   \Omega_M&=\big\{\boldsymbol{M}:\boldsymbol{M}\geq\mathbf{0}\big\}\\
  \Omega_{{a}}&=\big\{\balpha:\balpha\geq\mathbf{0};~{a}_1+{a}_2+\cdots+{a}_R = 1\big\}
\end{align}
where $\cdot\geq\mathbf{0}$ denotes elementwise nonnegativity of the argument matrix.% \cblue{matrix and once again all the symbols used in this article are in congruence with Table~\ref{tab_notation}}.

An observed spectrum $\by$ can be described by the following general mixture mechanism: 
\begin{equation}
           \label{eq_generalmodel}
           \by = \mathcal{F}^\star(\bM, \balpha) + \beps
\end{equation}
where $\mathcal{F}^\star$ defines the inherent photon interaction mechanism with the endmembers parameterized with their abundance fractions. In a nutshell, spectral unmixing aims at determining the endmembers and their respective abundances by considering the inverse process of $(\mathcal{F^\star})^{ -1}: \by \rightarrow \{\hat{\bM}, \hat{\balpha}\}$ under the constraints defined by $\Omega_M$ and $\Omega_{{a}}$, provided that this inverse is known or can be estimated. From the perspective of mathematical optimization, the general form of the unmixing problem can be formulated in different terms depending on the known information and priors.

%\subsection{{Problem formulations of the classical unmixing methods}}
\noindent \textbf{Problem formulation-I:} Under the assumption that the mixture mechanism is defined by an explicit $\mathcal{F}$, the optimization formulation of the unmixing problem is given by:
\boxedeq{eq_Prob1}{
           \begin{split}
            \big\{\hat{\bM}, \{\hat{\balpha}_i\}_{i=1}^N \big\}= &\mathop{\rm argmin}_{\bM, \{\balpha_i\}_{i=1}^N} \sum_{i=1}^N \mathcal{L}\big(\by_i,\hat{\by}_i\big) + \mathcal{R}\big(\bM, \{\balpha_i\}_{i=1}^N\big)\\
           & \qquad {\rm with } \quad   \hat{\by}_i = {\mathcal{F}}(\bM, \balpha_i) \\
           & \qquad\;\; {\rm s.t. } \quad   \bM\in\Omega_M, \, {\rm and \;} \balpha_i\in\Omega_a \; \text{for }\, \forall \, i
           \end{split} 
}
where
\begin{itemize}
     \item $\mathcal{L}(\cdot, \cdot)$ is a loss function defining the dissimilarity between its two arguments,
     \item $\mathcal{R}$ represents the regularizer that incorporates the prior information on endmembers $\bM$ and abundances $\{\balpha_i\}_{i=1}^N$.
\end{itemize} 

Extensive research has been conducted to derive explicit mathematical formulations for describing mixture mechanisms. These physics-based models are constructed with different assumptions regarding how photons interact with scene materials.
The works in~\cite{dobigeon2013nonlinear,Heylen2014A} provide thorough reviews on physical mixture models and associated unmixing methods. Some typical models are presented as examples below:

\noindent\textbf{{LMM.}} In the linear mixture model (LMM), the macroscopically pure components are assumed to be homogeneously distributed in separate patches within view field. Each photon can only interact with one pure material before reaching the observer, without being affected by other materials. The LMM is expressed as follows:
\begin{equation}\label{eq_lmm}
\begin{split}
\boldsymbol{y}&= \bm_1{a}_1 + \bm_2{a}_2 + \cdots +\bm_R{a}_R + \beps \\ &= \boldsymbol{M}\balpha+\beps,
\end{split}
\end{equation}
where the observed spectrum $\boldsymbol{y}$ is a linear combination of $\bm_i$, the spectrum of the $i$-th material and weighted by the associated abundance $a_i$. An additive noise is denoted by $\beps$. A general form of the LMM considers different endmembers in each pixel due to, e.g., spectral variability, and is formulated as:
\begin{equation}\label{eq_lmm_sp}
\begin{split}
\boldsymbol{y}_i&= \boldsymbol{M}_i\balpha_i+\beps_i,
\end{split}
\end{equation}
where $\boldsymbol{M}_i$, having the same structure as $\boldsymbol{M}$ defined in Table~\ref{tab_notation}, is the endmember matrix at the $i$-th pixel. 
%The LMM is easy to use and has a clear physical interpretation, but it cannot describe complex scenes.

\noindent\textbf{{Bilinear model.}} Bilinear models{~\cite{Halimi2011}} generalize the LMM by introducing second-order reflection signatures to capture, e.g.,  photon scatterings caused by multiple vegetation layers in agricultural scenarios. The mixture model is described by:
\begin{equation}\label{eq_bilinear}
  \boldsymbol{y}=\boldsymbol{M}\balpha+\sum_{i=1}^{R}\sum_{j=1}^{R}\gamma_{i,j}
  \boldsymbol{m}_i\odot\boldsymbol{m}_j+\beps,
\end{equation}
where $\odot$ denotes the Hadamard product, and {$\gamma_{i,j}$} stands for the nonlinear contribution of the $i$-th and $j$-th endmembers. The expression of $\gamma_{i,j}$ allows us to define different bilinear models, e.g., $\gamma_{i,j} = \beta_{i,j} a_i a_j$ yields the generalized bilinear model in~\cite{Halimi2011}, where $\beta_{i,j}$ controls the interaction strength between the $i$-th and $j$-th endmembers.
%Bilinear models have intuitive physical interpretations and can be easily extended to higher order reflections. However they have limited nonlinearity modeling capacity and their complexity dramatically increases when higher-order interactions are considered.

\noindent\textbf{{Hapke model.}} The Hapke model has been developed to characterize intimate mixtures with complex photon interactions~\cite{hapke2012theory}. The full Hapke model is complex and requires a number of parameters that are generally unavailable in real-world scenes. Under certain reasonable assumptions and simplifications such as spherical particles and isotropic scatters, the relationship between the bidirectional reflectance $\boldsymbol{y}$ and single-scattering albedo (SSA) $\boldsymbol{w}$ becomes:
\begin{equation}\label{eq_hapke}
\boldsymbol{y} =\mathcal{S}(\bw)=\frac{\boldsymbol{w}}{(1+2 \mu \sqrt{1-\boldsymbol{w}})\left(1+2 \mu_{0} \sqrt{1-\boldsymbol{w}}\right)}
\end{equation}
where the division is the element-wise operator, $\mu_0$ and $\mu$ are the cosines of the incoming and outgoing radiation angles, respectively. In spite of being nonlinear in terms of reflectance, the intimate mixture model is linear in the SSA domain and can be expressed as follows:
% researches demonstrate that it behaves linear mixture in the SSA domain, i.e.,
\begin{equation}
\boldsymbol{w}=\sum_{i=1}^{R} \boldsymbol{w}_{i} {{a}}_{i}.
\end{equation}

%Most unmixing problems with physics-based models fall into Problem formulation-I. 

{Unmixing problems with physics-based models frequently fall under Problem formulation-I when $\mathcal{F}$ is explicitly defined. Below are two examples of linear unmixing with known and unknown endmembers, respectively.}

\noindent\texttt{Example \theExpNo\stepcounter{ExpNo} (The FCLS unmixing)}: Under the linear model assumption $\mathcal{F}(\bM, \balpha)=\bM\balpha$ with known $\bM$, considering independent pixels and using the squared $\ell_2$-norm error $\mathcal{L}(\by_i, \hat{\by}_i) = \|\by_i-\hat{\by}_i\|^2$,  problem~\eqref{eq_Prob1} reduces to the Fully Constrained Least Square (FCLS) problem~{\cite{heylen2011fully}}: 
\begin{equation}
\label{eq_FCLS}
    \begin{split}
  \hat{\balpha}    = &\mathop{\rm argmin}_{\balpha} \|\by-\bM\balpha\|^2\\
          & \qquad\;\; {\rm s.t. } \quad   \balpha\in\Omega_\alpha.
  \end{split}
\end{equation}
\noindent\texttt{Example \theExpNo\stepcounter{ExpNo} (The regularized NMF unmixing)}: Under linear model assumption $\mathcal{F}(\bM, \balpha_i)=\bM\balpha_i$ for $i = 1, \cdots, N$, and using $\mathcal{L}(\by_i, \hat{\by}_i) = \|\by_i-\hat{\by}_i\|^2$ with a regularization term, problem~\eqref{eq_Prob1} becomes:
\begin{equation}
\label{eq_FCLS_NMF}
    \begin{split}
  \big\{\hat{\bM}, \{\hat{\balpha}_i\}_{i=1}^N \big\}    = &\mathop{\rm argmin}_{\bM, \{\balpha_i\}_{i=1}^N} \|\bY-\bM\bA\|_F^2 +  \mathcal{R}\big(\bM, \{\balpha_i\}_{i=1}^N\big)\\
          & \qquad\;\; {\rm s.t. } \quad   \bM\in\Omega_M, \, {\rm and \;}  \balpha_i\in\Omega_{{a}},
  \end{split}
\end{equation}
where we notice that $\|\bY-\bM\bA\|_F^2  = \sum_{i=1}^{N}\|\by_i -\bM\balpha_i\|^2$. {The design of regularizers $\mathcal{R}$ for enhancing the unmixing performance has been extensively studied in the literature. For example, the minimum volume-based term is usually introduced as a geometrical constraint on the endmember estimation~\cite{miao2007endmember}. Total-variation, sparsity and low-rankness are usually used to regularize the abundance estimation~\cite{peng2021low}.}

Though the physics-based model and unmixing with Problem formulation-I have been widely used, it is clear that the mixture mechanism may not be explicitly known or accurately assumed, and the existing physics-based models have limited nonlinearity modeling capacity. Simultaneously learning the function $\mathcal{F}$ from data while determining the endmembers and abundances becomes a remedy to this problem.

\noindent \textbf{Problem formulation-II: }  Under the assumption that the mixture mechanism $\mathcal{F}$ is unknown or partially unknown, the optimization formulation that learns $\mathcal{F}$ from the data can be formulated as:

\boxedeq{eq_Prob2}{
           \begin{split}
            \big\{\hat{\bM}, \{\hat{\balpha}_i\}_{i=1}^N, \hat{\mathcal{F}}\big\} = &\mathop{\rm argmin}_{\bM, \{\balpha_i\}_{i=1}^N, \mathcal{F}\in\mathcal{H}} \quad \sum_{i=1}^N \mathcal{L}\big(\by_i,\hat{\by}_i\big) + \mathcal{R}\big(\bM, \{\balpha_i\}_{i=1}^N\big)\\
           & \qquad {\rm with } \quad   \hat{\by}_i = {\mathcal{F}}\big(\bM, \balpha_i\big) \\
               & \qquad\;\; {\rm s.t. } \quad   \bM\in\Omega_M, \, {\rm and \;}   \balpha_i\in\Omega_{{a}}
           \end{split} 
}
with $\mathcal{H}$ being a functional space where $\mathcal{F}$ lies.  Typical nonlinear unmixing problems with conventional machine learning techniques often fall into this formulation{~\cite{li2019kernel,chen2013nonlinear_sp}}.

\begin{center}
\setlength{\fboxrule}{1pt}
\setlength{\fboxsep}{10pt}
\fcolorbox{bdblue}{bkblue}{
\begin{minipage} [t] {0.9\textwidth} 
Problem formulations-I and -II imply that the three steps outlined below are essential for achieving good unmixing performance:
\begin{itemize}
       \item Determining a mixture model $\mathcal{F}$ that matches the photon interaction mechanism;
       \item Defining a loss function that characterizes the dissimilarity between the reconstructed and observed spectra;
       \item Constructing a regularizer that captures the prior information on the variables.
\end{itemize}
\end{minipage}
}
\end{center}
\medskip

Learning $\mathcal{F}$ from data opens the possibility of determining mixture mechanisms that are difficult to explain using explicit descriptions and expressions. However, this strategy may fail if $\mathcal{F}$ is not sufficiently parameterized and jointly constrained with physics-based models. Some researchers have specified several forms for $\mathcal{F}$ to make it more interpretable. Except for the trivial linear form, the additive nonlinear model and post-nonlinear model are two major categories that have been extensively investigated.

\noindent\textbf{Additive nonlinear models.} Models in this category consider that the mixture consists of a linear mixture with an additive nonlinear component involving the endmembers and their abundances, that is,
\begin{equation}\label{eq_ma}
     \by = \bM\balpha + \mathcal{F}_{\rm add}(\bM, \balpha) + \beps.
\end{equation}
Several subcategories, depending on specific forms of $\mathcal{F}_{\rm add}$, have been investigated to make the problem more tractable, including:
\begin{itemize}
      \item The linear mixture/nonlinear fluctuation model: This model assumes that the mixture consists of two components, namely, a linear mixture component and a nonlinear fluctuation depending on the interaction between the endmembers, that is,
\begin{equation}
      \label{eq_anm}
           \by = \bM\balpha + \mathcal{F}_{\rm add}(\bm_1, \cdots, \bm_R) + \beps.
\end{equation}        
Though being useful and attractive, this model has some limitations: the additive nonlinear fluctuation function of this model appears to be independent of the abundances, and all endmembers contribute uniformly to the nonlinear component of the model. 
      \item The generalized linear-mixture/nonlinear-fluctuation model: This form suggests that the contribution of each material to the nonlinear interaction is proportional to their respective abundances, that is,
\begin{equation}
     \label{eq_ganm}
     \by = \bM\balpha + \mathcal{F}_{\rm add}({a}_1\bm_1, \cdots, {a}_R\bm_R) + \beps.
\end{equation}
This assumption appears reasonable given that a material with insignificant abundance will contribute minimally to both the linear and nonlinear mixing components in $\by$, and vice versa.
\end{itemize}

\noindent\textbf{Post-nonlinear models.}  Models in this category consider that a spectrum is produced by distorting the LMM with a nonlinear function, that is,
\begin{equation}
      \label{eq_pnm}
     \by = \mathcal{F}_{\rm post}(\bM \balpha) + \beps.
\end{equation}
A more specific form is constructed by setting $\mathcal{F}_{\rm post} = I+\mathcal{F}_{\rm add}$, leading to the additive post-nonlinear model:
\begin{equation}
     \label{eq_apnm}
     \by = \bM\balpha + \mathcal{F}_{\rm add}(\bM\balpha) + \beps,
\end{equation}
which can also be viewed as a case of~\eqref{eq_ganm} by setting $\mathcal{F}_{\rm add}({a}_1\bm_1, \cdots, {a}_R\bm_R) = \mathcal{F}_{\rm add}({a}_1\bm_1+\cdots+{a}_R\bm_R)$.

\begin{figure*}[!t]
  \centering
  % Requires \usepackage{graphicx}
  \includegraphics[width=18cm]{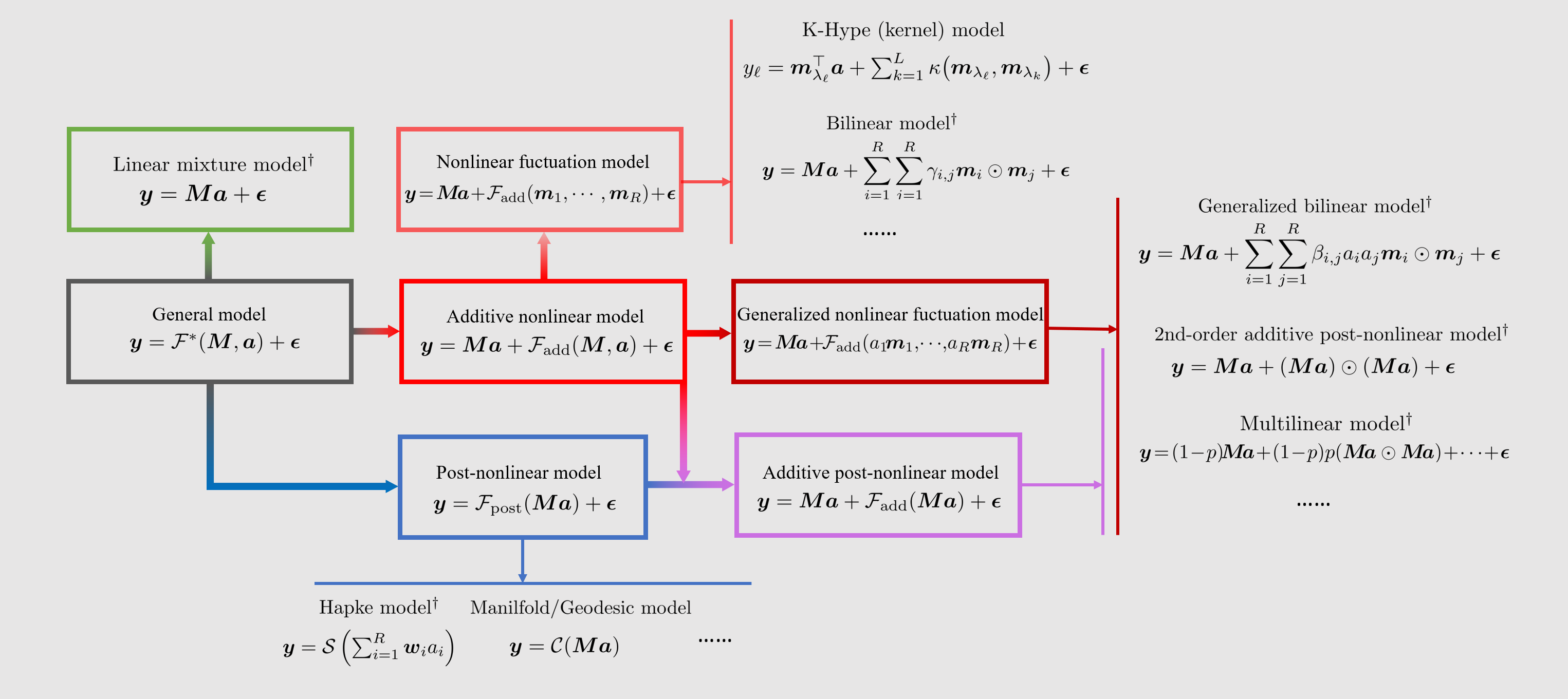}
  \vspace{-5mm}
  \caption{Hierarchical structures of general mixture models and physics-based models. The arrows start from general models to specific models. Physics-based models are marked with $\dag$.}
  \label{fig_modelrelation}
    \vspace{-3mm}
\end{figure*}

%\noindent\textbf{Endmember nonlinearity models.}  {To check again]}
%\begin{equation}
%     \label{eq_enm}
%     \by = \mathcal{F}_{\rm end}(\bM)\balpha + \beps.
%\end{equation}
\smallskip
Fig.~\ref{fig_modelrelation} shows the hierarchical structures of the above categories and how they cover typical physics-based mixture models. The examples below illustrate the relationship between Problem formulation-II and two typical conventional nonlinear unmixing techniques.

\noindent\texttt{Example \theExpNo\stepcounter{ExpNo} (K-Hype and regularized K-Hype unmixing):} The derivation of the K-Hype algorithm is based on a linear mixture/nonlinear fluctuation model of the form~\eqref{eq_anm}. The nonlinear fluctuation function $\mathcal{F}_{\rm add}$ is constrained to be a member of Reproducing Kernel Hilbert Space (RKHS) {$\mathcal{H}_{\rm RKHS}$} and expressed with kernels.
Considering independent pixels and assuming that $\bM$ is known, the optimization problem is given by:
\begin{equation}
\begin{split}
      \{\hat{\balpha}, \hat{\mathcal{F}}_{\rm add}\} = &\mathop{\rm argmin}_{\balpha, \mathcal{F}_{\rm add}\in\mathcal{H}_{\rm RKHS}} \frac{1}{2\mu}\sum_{\ell=1}^L e_\ell^2 +  \frac{1}{2}\Big(\|\mathcal{F}_{\rm add}\|^2 + \|\balpha\|^2\Big)
\\
  & \qquad {\rm with } \quad e_\ell = y_\ell - \bm_{\lambda_\ell}^\top \balpha - \mathcal{F}_{\rm add}(\bm_{\lambda_\ell}) \\
               & \qquad\;\; {\rm s.t. } \quad   \balpha\in\Omega_a
\end{split} 
\end{equation}
where $\bm_{\lambda_\ell}$ denotes the $\ell$-th row of $\bM$, i.e.,  all reflectance at band $\lambda_\ell$.
This problem can be solved using the dual method, which yields an explicit expression relating $\hat{\balpha}$ and the dual variables, and expands $\hat{\mathcal{F}}_{\rm add}$ as a sequence of kernels parameterized by $\{\bm_{\lambda_\ell}\}_{\ell=1}^L$. The K-Hype formulation, along with its sparsity-promotion variant~\cite{li2019kernel}, and its spatially regularized variant in~\cite{chen2013nonlinear_sp}, clearly fall into the form of Problem {formulation}-II.

\noindent\texttt{Example \theExpNo\stepcounter{ExpNo} ({Unmixing with manifold learning/geodesic distance)}:} The geodesic distance can be calculated to conduct data unmixing on a manifold rather than in a linear space. This process considers a nonlinear but continuous bijective mapping $\mathcal{C}$ between the linear space of abundance coefficients and a manifold in the spectral space:
\begin{equation}
\by = \mathcal{C}(\bM\balpha)+\beps,
\end{equation}
which is of the form~\eqref{eq_pnm} as shown in~\cite{heylen2010non, heylen2014distance}. {More specifically, defining a K-Nearest Neighbor (KNN) graph with the data set, the geodesic distance can be approximated as the length of the shortest path between two points. The endmembers $\{\bm_{i}\}_{i=1}^R$ span the simplex with the largest volume on the data manifold, and the abundance coefficients of each data to be unmixed can be expressed as volume ratios in this manifold.}
% \begin{equation}
%       \balpha_i = \frac{\bV_{(\bm_{1} \cdots \bm_{i-1}, \boldsymbol{y}_i, \bm_{i+1}, \cdots,  \bm_{R})}}{\bV_{(\bm_{1} \cdots \bm_{R})}}
% \end{equation}
%where the volume $\bV$ is the inter-vertex distance using Cayley-Menger determinant. 
{The manifold formulation also falls into Problem {formulation}-II as $\mathcal{C}$ is learnt from data.}

{Problem formulations-I and -II cover physics-based methods or conventional machine learning techniques with handcrafted regularizers. This article focuses more on the integration of physics-based methods with recent machine learning techniques powered by DNNs and other data-driven techniques. Problem formulation-III encompasses these new methods by introducing additional hierarchical structures on endmembers and abundances.}

\noindent \textbf{Problem formulation-III: } Under the assumption that the mixture mechanism $\mathcal{F}$ is unknown or partially unknown, the optimization formulation that i) jointly learns $\mathcal{F}$ from data, and that ii) considers hierarchical reparameterization of the endmembers and abundances, can be formulated as:
\boxedeq{eq_Prob3}{
           \begin{split}
            \big\{\hat{\bM}, \bTheta_{\bM}^\star,\{\hat{\balpha}_i\}_{i=1}^N, {\bTheta_{\ba}^\star}, \hat{\mathcal{F}}\big\} = &\mathop{\rm argmin}_{\bM, \bTheta_{\bM}, \{\balpha_i\}_{i=1}^N, \bTheta_{\ba}, \mathcal{F}\in\mathcal{H}} \quad \sum_{i=1}^N \mathcal{L}\big(\by_i,\hat{\by}_i\big) + \mathcal{R}\big(\bM, \{\balpha_i\}_{i=1}^N\big)\\
           & \qquad {\rm with } \quad   \hat{\by}_i = {\mathcal{F}}\big(\bM(\bTheta_{\bM}), \balpha_i(\bTheta_{\ba})\big) \\
            & \qquad\;\; {\rm s.t. } \quad   \bM\in\Omega_M, \, {\rm and \;}   \balpha_i\in\Omega_{{a}}
           \end{split}
}
where $\bTheta_{\bM}$ and $\bTheta_{\ba}$ are parameters that allow us to determine the endmember matrix $\bM$ and the abundances $\{\balpha_i\}_{i=1}^N$, respectively, and {$\mathcal{H}$ denotes a given function space.}  For example, in recent {DNN} structures, $\bTheta_{\ba}$ can represent the encoder in an autoencoder unmixing framework, and $\bTheta_{\bM}$ can represent the decoder part that generates the endmembers. Recent physically inspired data-driven methods will be reviewed in the following three sections. Their connections to Problem formulation-III will be also established.
% In contrast to unmixing methods based on conventional machine learning techniques that mostly focus on learning from data,
{In contrast to conventional machine learning based unmixing methods that mostly focus on learning  mixture models from data, recent data-driven methods show their superiority in improving a variety of aspects including model design, prior capturing, and loss function design.}

\section{Integrating of physics-based models in {DNN} design} 

\begin{figure*}[!t]
  \centering
  % Requires \usepackage{graphicx}
  \includegraphics[width=18cm]{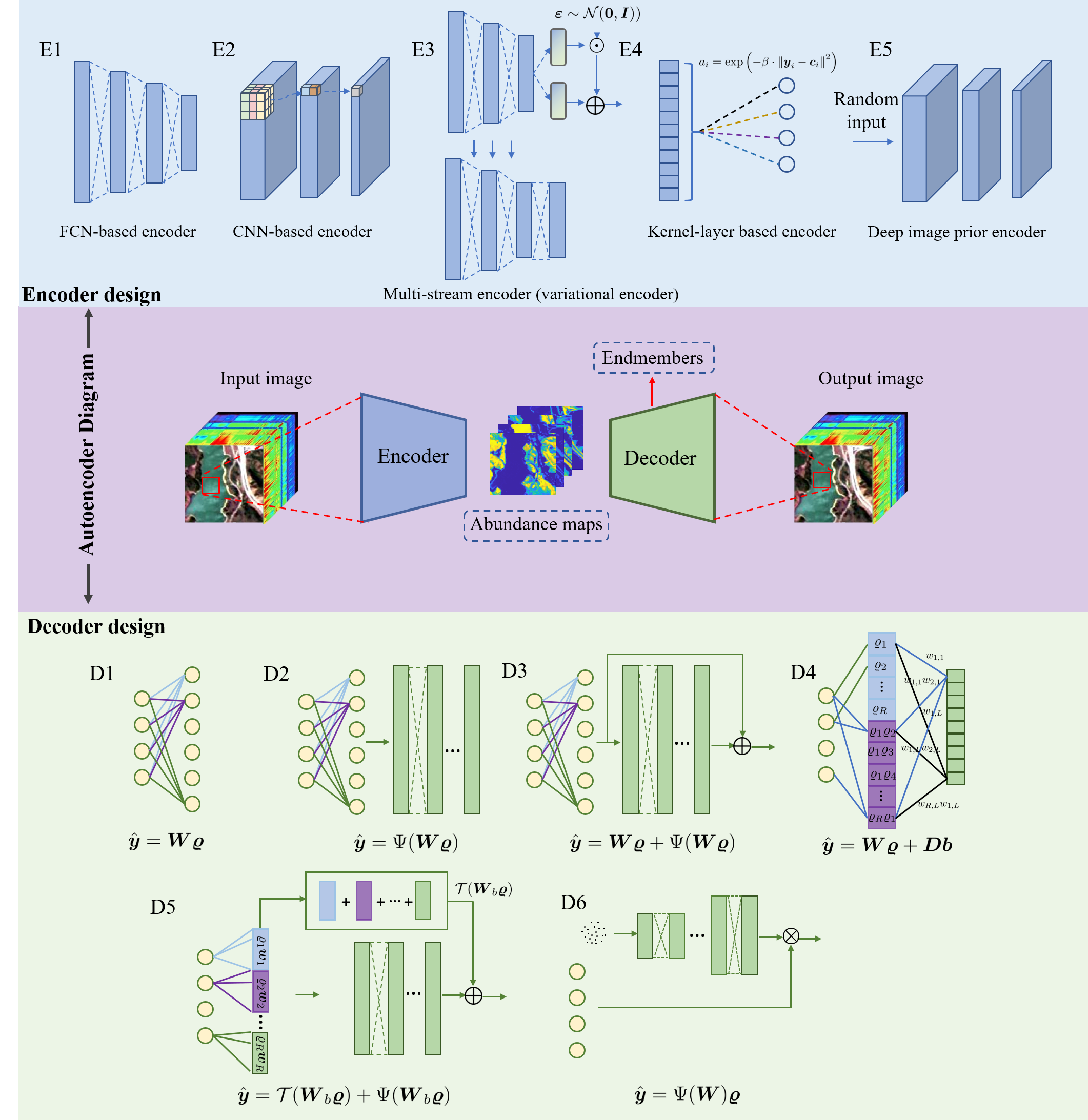}
    \vspace{-5mm}
  \caption{{This figure illustrates the structure of an autoencoder for hyperspectral unmixing and a variety of selections of encoder and decoder designs.}}
  \label{fig_AECmodels}
    \vspace{-3mm}
\end{figure*}

In practice, it appears natural to model $\mathcal{F}^\star$ in~\eqref{eq_generalmodel} using {DNNs} since they have enhanced modeling capacity compared to conventional nonlinear modeling techniques. Instead of using {DNNs} as black boxes, elegant designs of network structures for modeling~\eqref{eq_ganm}--\eqref{eq_apnm} are indispensable for preserving physical interpretation and then to facilitate the extraction of abundances and endmembers. Deep autoencoder networks are considered as the most appropriate ones and are intensively studied for this purpose. Essentially, an autoencoder consists of an encoder network and a decoder network. The encoder part seeks for a low dimensional representation (e.g., the abundances) to reconstruct the input image. Given an input $\by$, 
\begin{equation}
      \boldsymbol{\varrho} = f_{\bTheta_{\rm enc}}(\by)
\end{equation}
where $f_{\bTheta_{\rm enc}}$ is the function representing the encoder with $\bTheta_{\rm enc}$ denoting all network parameters. The decoder part reverses the functionality of the encoder by decompressing the hidden representation vector to reconstruct the original input data, i.e.: %with the corresponding base (e.g., the endmember). 
\begin{equation}
      \hat{\boldsymbol{y}} = f_{\bTheta_{\rm dec}}(\boldsymbol{\varrho})
\end{equation}
where $f_{\bTheta_{\rm dec}}$ is the function representing the decoder with $\bTheta_{\rm dec}$ denoting all network parameters. For instance, an autoencoder can be designed to consider $\boldsymbol{\varrho}$ as an estimate of the abundances $\balpha$, and determine endmembers and mixture mechanism from the decoder network. Then, identifications ${\bTheta_{\ba}}\leftarrow\bTheta_{\rm enc}$ and $\mathcal{F}\leftarrow f_{\bTheta_{\rm dec}}$ can be considered to fit these structures into Problem formulation-III. In the following, rather than simply presenting the original autoencoders introduced in existing works as a whole, we show the designs of encoders and decoders separately which then demonstrates their extra flexibility and the potential of combining different designs. The basic structure of an autoencoder and typical encoder and decoder designs for unmixing are summarized in Fig. \ref{fig_AECmodels}.  We start with decoders that are in fact associated with mixture models. Then, we proceed to the encoders that implicitly benefit from data priors.

\subsection{Design of structured decoders to learn models from data}
%We start with a single linear layer for the decoder, and move forward step-by-step to show how mixture models learnt by an autoencoder depend on the decoder.
Beginning with the decoder of a single linear layer, we progressively demonstrate that the mixture models learned by an autoencoder are actually dependent on the decoder.

\subsubsection{{Linear decoder}}
The relationship between the input and output of a fully connected linear layer decoder is as follows:
\begin{equation}\label{eq_d_linear}
     \hat{\by} = f_{\bTheta_{\rm dec}}(\boldsymbol{\varrho})=\boldsymbol{W} \boldsymbol{\varrho}.
\end{equation}
As shown in D1 of Fig.~\ref{fig_AECmodels}, this form is consistent with the LMM defined in~\eqref{eq_lmm}. Consequently, the learning process of the autoencoder, which minimizes the gap between $\hat{\by}$ and $\by$, leads to $\boldsymbol{\varrho}$ as an estimate of $\balpha$, and $\boldsymbol{W}$ as an estimate of the endmembers matrix $\bM$. Despite the fact that the linear unmixing problem has been extensively studied with various classes of methods~\cite{ozkan2018end,qu2018udas,su2019daen,palsson2018hyperspectral,palsson2020convolutional,khajehrayeni2020hyperspectral}, this structure motivates the other sophisticated designs that learn structured models from the data.

\subsubsection{{Post-nonlinear decoder}}
With the linear decoder at hand, one can connect it to additional nonlinear layers in order to create the post-nonlinear structure:
\begin{equation}
     \hat{\by} = f_{\bTheta_{\rm dec}}(\boldsymbol{\varrho})= \Psi(\boldsymbol{W} \boldsymbol{\varrho})
\end{equation}
 where $\Psi$ denotes an implicit mapping function represented by the subsequent network layers with nonlinear activation functions. Parameters $\bTheta_{\rm dec}$ can be determined by training the autoencoder so that the nonlinear function is learnt from the data. Compared with the post-nonlinear mixture model defined in~\eqref{eq_pnm}, $\Psi$ is learnt to mimic the generative model $\mathcal{F}_{\rm post}$, and the endmembers are obtained from the weights of the first layer of the decoder~\cite{NAE2019}. The  structure  that  forms  this  process is shown by D2 in Fig. \ref{fig_AECmodels}.
 
\subsubsection{{Additive nonlinear decoders}} Instead of adding subsequential nonlinear layers, a network branch can be added in parallel to the linear layer for constructing additive nonlinear structures. However, the parameters coupled in the linear and the nonlinear components should be carefully addressed.
\begin{itemize}
    \item Additive post-nonlinear decoder: Passing the linear component through a nonlinear network and adding the output to the linear component, leads to the following result: 
    \begin{equation}\label{eq_d_add_p}
     \hat{\by} = \bW\boldsymbol{\varrho} + \Psi( \bW\boldsymbol{\varrho}). 
\end{equation}
    The structure that conducts this process is depicted in D3 of Fig. \ref{fig_AECmodels}. Comparing the output of this decoder to the model in~\eqref{eq_apnm}, the endmember can be estimated from the network weights of the linear component, and $\Psi$ mimics $\mathcal{F}_{\rm add}$ in~\eqref{eq_apnm} once the network is learnt~\cite{zhao2021lstm}.
    \item Generalized additive nonlinear decoder: Constructing a decoder that mimics the generalized linear mixture/nonlinear fluctuation model~\eqref{eq_ganm} requires more specific designs. First, a block diagonal matrix $\bW_b = \text{bdiag}\{\bw_1, \cdots, \bw_R\}$ is considered for the weights of the first linear layer, which is then sparsely connected. Consequently, the output of this layer is a collection of $\bw_i$ weighted by $\varrho_i$, i.e.:
    \begin{equation}
        \bW_b \boldsymbol{\varrho} = [\varrho_1\bw_1^\top, \varrho_2\bw_2^\top, \cdots, \varrho_R\bw_R^\top]^\top.
    \end{equation}
    On one hand, the linearly-mixed component of the model can be obtained by adding all $R$-spaced entries of $\bW_b \boldsymbol{\varrho}$ such that:
    \begin{equation}
        \mathcal{T}(\bW_b \boldsymbol{\varrho}) = \bW\boldsymbol{\varrho} 
    \end{equation}
    with $\mathcal{T}: \mathbb{R}^{LR}\mapsto\mathbb{R}^{L} $ defined as $\mathcal{T}(\bx) = \left[\sum_{i=1}^R x_{R(i-1)+1}, \sum_{i=1}^R x_{R(i-1)+2}, \cdots, \sum_{i=1}^R x_{R(i-1)+L} \right]^\top$. On the other hand, the weighted nonlinear fluctuation component is obtained by feeding $\bW_b \boldsymbol{\varrho}$ to a nonlinear network representing $\Psi$. Finally, the sum of these two components:
    \begin{equation}\label{eq_d_add_g}
        \hat{\by} = \mathcal{T}(\bW_b \boldsymbol{\varrho}) + \Psi(\bW_b \boldsymbol{\varrho})
    \end{equation}
    is consistent with the model in~\eqref{eq_ganm}. The endmembers are estimated by $\bW$ and the nonlinear mixing function $\mathcal{F}_{\rm add}$ is estimated by $\Psi$ from training data~\cite{zhao2021hyperspectral}. The network structure to achieve this design is depicted in D5 of Fig. \ref{fig_AECmodels}. 
    \item {Bilinear decoder}: Designing a decoder to mimic the bilinear model defined in \eqref{eq_bilinear} leads to an autoencoder  formulated as:
    \begin{equation}\label{eq_d_bilinear}
    \hat{\by}= \bW\boldsymbol{\varrho}+{\boldsymbol{D}\boldsymbol{b}},
    \end{equation}
    where {$\boldsymbol{D}=[\boldsymbol{d}_1,\boldsymbol{d}_2,\cdots,\boldsymbol{d}_c]\in\mathbb{R}^{L\times c}$}
    is a cross-product matrix involving all pairs of endmembers, i.e., {$\boldsymbol{d}_{i,j}=\bw_{i}\odot\bw_{j}$} with $c=\frac{R(R-1)}{2}$, and {$\boldsymbol{b}=[b_1,b_2,\cdots,b_c]\in\mathbb{R}^{c}$} denotes the interaction abundance matrix. The structure of this model can be seen from D4 in Fig. \ref{fig_AECmodels}. Referring to \eqref{eq_bilinear}, the weights of the decoder provide estimates of the endmembers and their cross-products. Vector {$\boldsymbol{b}$} provides estimates of the cross-products of all pairs of abundances ${\boldsymbol{b}}=[\varrho_1\varrho_2,\varrho_2\varrho_3,\cdots, \varrho_R\varrho_1]$ or the latent features learnt from the data~\cite{su2020deep,shahid2021unsupervised}.
\end{itemize}

\subsubsection{{Endmember generating decoder}} Instead of being extracted from the weights of a decoder, the endmembers can also be estimated from a sub-network, that is,
\begin{equation}\label{eq_d_endnon}
     \hat{\by} = f_{\Theta_{\rm end}}( \bz)\boldsymbol{\varrho}=\Psi(\bz)\boldsymbol{\varrho}
\end{equation}
where $f_{\Theta_{\rm end}}$ represents the endmember generative model, with $\boldsymbol {\Theta}_{\rm end}$ denoting the network parameters. This sub-network can be a regular or variational autoencoder that maps pixels to endmembers~\cite{jin2021tanet,shi2021probabilistic} or a pretrained decoder to generate the endmembers from a latent vector~\cite{borsoi2019deep}. Variable $\boldsymbol{z}$ denotes the input of this sub-network, which can be either pixels, candidate endmembers or latent representations. This decoder is also a powerful tool to cope with spectral variability~\cite{shi2021probabilistic,borsoi2019deep}. Compared with Problem formulation-III, we have the identification $\bTheta_{\bM}\leftarrow\bTheta_{\rm end}$, leading to a hierarchical structure $\bM(\bTheta_{\bM})$.

\subsection{Design of encoders to integrate priors from data}

The encoder part is designed to map the input spectra into a latent representation space, which is often considered as the {abundance} space.  The encoder implicitly acts as a regularizer relating abundances and spectra, and imposes a hierarchical structure on $\balpha$ with parameters $\bTheta_{\balpha}$, i.e., of the form $\balpha(\bTheta_{\balpha})$.  The encoder part has been designed in the following ways in the literature.
%Similarly, the endmebmers $\bM$ are also considered as output by some encoders, leading to a hierarchical structure $\bM(\bTheta_{\bM})$.

%y relying the decoder that defines the mixture model $\mathcal{F}$ leads to Problem formulation-II. However the solution to the problem can be non-stable since the associated problems can be highly nonconvex the solution. Using an encoder relating the spectra and the abundances actually implicitly plays a role of regularization, and imposes a hierarchical structure on $\balpha$ with parameters $\bTheta_{\balpha}$, i.e., in form of $\balpha(\bTheta_{\balpha)}$. Similarly, the endmebmers $\bM$ are also considered as output by some encoders, leading to a hierarchical structure $\bM(\bTheta_{\bM})$.

%The encoder part has been designed as follows in the literature:

\subsubsection{{FCN-based encoders}} Early works~\cite{palsson2018hyperspectral,ozkan2018end,qu2018udas,su2019daen} employ fully connected layers to construct the encoder. The latent representation of the $k$-th layer is given by
\begin{equation}
     \begin{split}
     &\bh_{k} = \sigma_k(\bU_k \bh_{k-1}  + \bb_{k}) \\
     & \text{with} \qquad \bh_{0} = \by
     \end{split}
\end{equation}
where $\sigma_k$ represents the nonlinear activation function, and $\bU_{k}$ and $\bb_k$ stand for the weights and bias of the $k$-th layer, respectively. The output of the final layer is denoted by $\boldsymbol{\varrho}$ and is connected to the decoder. Some variants have also been designed based on this structure. For example, in~\cite{ozkan2018end,qu2018udas}, a sparse encoder is used to enforce abundance sparsity. The work~\cite{qu2018udas} applies a denoising encoder to address the problem of noisy data and outliers in hyperspectral data.

% \cblue{The sparsity of the encoder output is enforced in~\cite{ozkan2018end,qu2018udas} by constraining the activation number of hidden units, especially the last layer of encoder. To solve the noise and outliers in hyperspectral data, the denoising encoder is introduced~\cite{qu2018udas}. The input data is manually perturbed and the network is trained to recover the original clean input. Through this structure the input pixel can be robust to small variation.}

\subsubsection{{CNN-based encoders}} Convolutional neural networks (CNNs) have also been used for designing the encoder~\cite{palsson2020convolutional,khajehrayeni2020hyperspectral,zhao2021hyperspectral}. Encoder inputs can be the entire image or overlapping patches. The 2D or 3D convolution operator operates on
3D image cubes $\mathbb{Y}_n$  as follows:
\begin{equation}\label{eq_cnn}
     \begin{split}
     &\mathbb{H}_{k} = \sigma_k(\mathbb{H}_{k-1} \circledast \mathbb{U}_{k}+\mathbb{B}_k) \\
     & {\rm with} \qquad \mathbb{H}_{0} = \mathbb{Y}_n,
     \end{split}
\end{equation}
% For the $n$-th pixel, the convolution operator is conducted on it and its neighborhood, which are defined as $\mathbb{y}$, and the output is the abundance estimate of this pixel.
% \begin{equation}\label{eq_cnn}
%      \begin{split}
%      &\bh_{k} = \sigma_k([\widetilde{\bh}_n]_{k-1} \circledast \bU_{k}+\bb_k) \\
%      & {\rm with} \qquad [\widetilde{\bh}_n]_{0} = [\widetilde{\by}_n] 
%      \end{split}
% \end{equation}
%
% The input of the encoder can be a patch of the image around pixel $n$, denoted by $[\bY]_n$, and the output is the abundance estimate of this pixel.
% \begin{equation}
%      \begin{split}
%      &\bH_{k} = \sigma_k(\bH_{k-1} \circledast \bU_{k} + \bb_k) \\
%      & {\rm with} \qquad \bH_{0} = [\bY]_n  {Not yet well formualted}
%      \end{split}
% \end{equation}
where $\mathbb{U}_{k}$ denotes a convolution kernel, and $\mathbb{B}_k$ represents the bias parameters.
As 2D convolutional kernels may cause local spectral distortion,  3D-CNN encoders are used in~\cite{khajehrayeni2020hyperspectral,zhao2021hyperspectral} to jointly learn spatial structures and local spectral features of 3D input images.
%{2D convolutional kernels analyze hyperspectral images causing to local spectral information distortion.
%3D-CNN~\cite{khajehrayeni2020hyperspectral,zhao2021hyperspectral} is utilized in the encoder to jointly learn the spatial structures and local spectral features from 3D input images.}

\subsubsection{{RNN encoders}}
Recurrent neural networks (RNNs), such as long short term memory (LSTM) networks, benefit from memory cells and are better suited for processing sequences. With their recurrent hidden states depending on previous steps, they can take full advantage of the sequential characteristics in hyperspectral data. The abundances are estimated with an LSTM to encode spectral correlation and band-to-band variability information~\cite{zhao2021lstm}.

% \noindent\textbf{\cblue{Sparse encoder:}}  The sparsity of the encoder output is enforced in~\cite{ozkan2018end,qu2018udas} by constraining the activation number of hidden units, especially the last layer of encoder. %As mixed pixels often contain a subset of endmembers, sparse encoder can be used to effectively enhance the abundance sparsity.

% \noindent\textbf{{Denoising encoder:}}
% To solve the noise and outliers in hyperspectral data, the denoising encoder is introduced~\cite{qu2018udas}. The input data is manually perturbed and the network is trained to recover the original clean input. Through this structure the input pixel can be robust to small variation.

\subsubsection{{Kernel-layer based encoder}}
Another class of approaches uses kernel layers to process hyperspectral data, and such layers can map the input pixels into a manifold and estimate the abundances according to a Riemannian metric.
For example, in~\cite{shahid2021unsupervised}, the kernel layer is defined as:
\begin{equation}
{a}_i=\exp\left(-\beta \cdot\left\|\by_{i}-\boldsymbol{c}_i\right\|^{2}\right)
\end{equation}
where $\beta$ is the kernel parameter, $\boldsymbol{c}_i$ denotes the centroid related to the $i$-th endmember. The smaller the distance between the mixed pixel and the endmember centroid is, the larger is the corresponding abundance. 

\subsubsection{{Variational encoders}} A variational encoder encodes each input pixel with a distribution in the latent space as opposed to a single vector. In practice, it is common to consider Gaussian distributions parameterized by their means and covariance matrices. The work~\cite{su2019daen} uses a variational encoder to infer abundances. In~\cite{shi2021probabilistic}, the probabilistic generative model is used to address endmember variability and to provide more accurate abundance and endmember estimates.
%
%is a probability model. It encodes the training pixel into a soft ellipsoidal region. Combining with the decoder, this architecture can generate new samples by sampling from the output of encoder from a parameterized probability function, usually $\mathcal{N}(\boldsymbol{0},\bI)$. The work~\cite{su2019daen} uses variational encoder to infer abundances. In~\cite{shi2021probabilistic}, the latent variables inferred by the encoder is considered as nonlinear representations of endmembers and used to cope with endmember variability issue.

\subsubsection{{Multi-branch encoders}} 
Some works use multi-branch encoders to capture different features by sharing weights or concatenating features. In~\cite{han2020deep}, an endmember guided encoder is used to transfer the endmember information into a parallel abundance estimate encoder. The work~\cite{hua2021dual} uses a dual branch encoder, where one branch consists of fully connected layers to extract spectral information, and the other branch consists of 2D convolution layers to capture spatial information. 

% Researchers also proposed some variants of encoder.
% In~\cite{}, a dual branch autoencoder network is involved to leverage spatial-contextual information. Specifically, One branch is formed by fully connected layers to extract spectral information, the other adopt 2D convolution layers to capture spatial information. Then the features of these two branches are fused for unmixing.
% The work~\cite{} use two stream autoencoders for spectral unmixing. In the first stream, superpixel segmentation strategy is used to add spatial correlation information.
% In~\cite{}, spatially aware filters built using radial basis function (RBF) kernels are  applied to design the encoder.  This method is robust in the presence of unresponsive pixels.

\noindent\texttt{Remark:} The ANC and the ASC are usually addressed at the last layer of the encoder either by using specific activation functions (e.g. ReLU, Sigmoid, absolute operator) to enforce ASC followed by a normalization layer to enforce ANC, or by using SoftMax to enforce ANC and ASC simultaneously. They can also be addressed approximatively via a penalty term in the loss function.

\noindent\texttt{Example \theExpNo\stepcounter{ExpNo} (The 3D-CNN based autoencoder network for additive nonlinearity unmixing)}: Considering the 3D-CNN based encoder (E2 in Fig.~\ref{fig_AECmodels}) and the generalized additive nonlinear decoder defined in \eqref{eq_d_add_g} (D5 in Fig.~\ref{fig_AECmodels}), leads to the algorithm proposed in~\cite{zhao2021hyperspectral}.

\noindent\texttt{Example \theExpNo\stepcounter{ExpNo} (A denoising autoencoder with sparsity for spectral unmixing)}: Considering the denoising and sparse encoder and the linear decoder defined in \eqref{eq_d_linear} (E1 and D1 in Fig.~\ref{fig_AECmodels}), leads to the method proposed in~\cite{qu2018udas}. %udas

\noindent\texttt{Example \theExpNo\stepcounter{ExpNo} (The  kernelization and cross product layer based autoencoder for bilinear \\ and post nonlinearity unmixing)}: Considering the kernel layer based encoder (E4 in Fig.~\ref{fig_AECmodels}), the bilinear nonlinear decoder defined in~\eqref{eq_d_bilinear} (D4 in Fig.~\ref{fig_AECmodels}) and the post nonlinear decoder defined in~\eqref{eq_d_add_p} (D3 in Fig.~\ref{fig_AECmodels}), corresponds to the method proposed in~\cite{shahid2021unsupervised}.

\noindent\texttt{Example \theExpNo\stepcounter{ExpNo} (The probabilistic generative model based autoencoder for spectral unmixing\\ with spectral variability)}: A two-stream autoencoder framework with a probabilistic generative model (E3 and D6 in Fig.~\ref{fig_AECmodels}) that copes with spectral variability, corresponds to the method proposed in~\cite{shi2021probabilistic}.

\noindent\texttt{Example \theExpNo\stepcounter{ExpNo} (Multitask based autoencoder network for bilinear spectral unmixing)}: 
The combination of two deep autoencoders with a multitask learning framework leads to the method in~\cite{su2020deep}.  One of the autoencoders models linear components and estimates the endmembers as well as their abundances, while the other models bilinear components defined in~\eqref{eq_d_bilinear} and estimates interaction abundances.

%Two deep autoencoders with a multitask learning framework, one to model linear components and estimate the endmembers and their abundances, the other to model bilinear components defined in and estimate interaction abundances, lead to the method proposed in~\cite{su2020deep}.

% and the bilinear model defined in~\eqref{eq_d_bilinear} (depicted in Fig.~\ref{fig_AECmodels} D4) leads to the method proposed in~\cite{su2020deep}.

In addition to the regular autoencoders that have been described thus far, there are also some variant autoencoder frameworks that have been proposed for spectral unmixing.  For instance, in~\cite{su2019daen}, a group of autoencoders is first applied in order to address issues regarding {outliers and generate a good initialization}.  Then, an autoencoder with a single-layer linear decoder is used to perform the process of unmixing. 
% The architecture proposed in \cite{mulsu2020deep} consists of two autoencoders that perform different tasks, one is to model linear components and estimate the endmembers and their abundances, the other is to model bilinear components and estimate interaction abundances. 
To improve the unmixing performance, two cascaded autoencoders are utilized in an end-to-end fashion to address cycle-consistency constraints in~\cite{gao2021cycu}.
 In~\cite{rasti2021undip}, the endmembers are extracted using a geometric endmember extraction method, and the abundances are estimated using a deep image prior based on a CNN (depicted by E5 in Fig.~\ref{fig_AECmodels}). In~\cite{qi2022sscu}, a two-stream network, with one stream for spectral information and the other for employing spatial information, is collaboratively learnt to make full use of the spectral and spatial information. The work~\cite{han2022multimodal} constructs an attention network to fuse the information of light detection and ranging (LiDAR) data into the unmixing process.

\smallskip

\begin{center}
\setlength{\fboxrule}{1pt}
\setlength{\fboxsep}{10pt}
\fcolorbox{bdblue}{bkblue}{
\begin{minipage} [t] {0.9\textwidth} 
Remarks on unmixing with deep autoencoder networks:
\begin{itemize}
       \item The underlying unmixing model used in an autoencoder network is determined by the structure of the decoder and independent of the encoder.
       \item The encoder implicitly incorporates the inference relation from the observed spectra to abundances, and can thus be considered as a prior for unmixing.
       \item Unmixing with an autoencoder falls into Problem formulation-III with $\mathcal{F}$ and $\bTheta_{\bM}$ associated with the corresponding parts of the decoder and $\bTheta_{\balpha}$ learnt by the encoder network.
\end{itemize}
\end{minipage}
}
\end{center}
%\medskip

% Besides, hyperspectral images are usually corrupted by noise or outliers, which may dramatically decrease the unmixing performance. To cope with this issue, denoising-oriented encoder architectures are derived. 
% In~\cite{su2019daen,1su2018stacked}, stacked autoencoders are utilized to address  noise and outlier. In the first step, several autoencoder based models are used to learn a good representation of the input and deal with the outliers. These autoencoders are all trained to minimize the reconstructed error, and the output of the previous one is the input of the latter. In the second step,  an autoencoder is conducted to perform unmixing. Denoising before the unmixing procedure may introduce additional reconstructed error before unmixing.
% The work~\cite{qu2018udas} proposes to integrate a denoising constraint to achieve
% the denoising capacity, in which the denoising reconstructed pixels from the marginalized denoising autoencoder (mDA) are used to regularize the ouput unmixing autoencoder. As mixed pixels often contain a subset of endmembers, sparsity-promoting regularizers can enhance the unmixing performance. There are mainly two strategies to enforce the sparsity of abundances, one is to enforce the activation of the last layer of encoder to be sparse, the other is to directly plug abundance sparsity constraints in the loss function.

%  \subsection{Integrating image priors learnt from data into physics-based inverse problems.} 
%  \subsection{Integrating inference algorithms in neural network design.}

\section{{Prior information learning with data-driven approaches}}

In addition to modeling the spectral mixture mechanism, DNNs are superior to conventional machine learning techniques
in extracting information from data priors.
It is intriguing to use DNNs to explicitly learn prior knowledge from data and use it with physical model-based inversion algorithms. {Instead of designing sophisticated regularizers,} the plug-and-play paradigm uses a denoiser with deep architectures to learn image prior and then incorporates the denoiser into an iterative optimization of the problem in~\eqref{eq_Prob1}. On the other hand, deep unrolling algorithms aim to unfold an iterative optimization of the problem in~\eqref{eq_Prob1} with a specific regularizer into a trainable end-to-end deep network.

\subsection{Incorporating data-driven priors with denoisers}
Benefiting from the variable splitting technique, the plug-and-play framework makes it possible to exploit the priors learnt from training data to solve various image restoration problems. State-of-the-art denoising algorithms, especially fast-speed and powerful CNN-based denoisers are usually plugged into this framework as the proximity operator that captures the intrinsic spatial and spectral structures of hyperspectral images.
%~\cite{dian2020regularizing,wang2020learning}
By leveraging the flexibility of model-based optimization and the powerful learning ability of CNN, this framework has shown its great potential to perform unmixing tasks. Among three problem formulations, Problem formulation-I is considered here for simplicty and illustration. In the plug-and-play framework, an auxiliary variable $\bZ$ is introduced in~\eqref{eq_Prob1} to replace $\bA$ in the regularization term, with an extra constraint $\bZ = \bA$ to ensure the equivalence of the optimization problems. Using the ADMM method results in the following iterative updates:
\begin{subequations}
	\begin{align}
	\label{eq:stepa_pnp} \big\{\bM^{(k+1)}, \bA^{(k+1)} \big\} = &\mathop{\rm argmin}_{\bM, \{\balpha_i\}_{i=1}^N} \sum_{i=1}^N \mathcal{L}\big(\by_i,\hat{\by}_i\big) + \frac{\rho}{2} \big\|\bA-\bZ^{(k)}+\bV^{(k)}\big\|_F^2 \qquad {\rm with } \quad   \hat{\by}_i = {\mathcal{F}}(\bM, \balpha_i) \\ \nonumber \qquad\;\;\;\; & {\rm s.t. } \quad   \bM\in\Omega_M, \, {\rm and \;} \balpha_i\in\Omega_{{a}}\\ 
	\label{eq:stepb_pnp} {\bZ}^{(k+1)}\; = &\;\ \mathop{\rm argmin}_{\bZ} \frac{\rho}{2} \big\|\bZ-(\bA^{(k+1)}+\bV^{(k)})\big\|_F^2 + \mathcal{R}(\bZ) \\
	\label{eq:stepc_pnp}
	\bV^{(k+1)}\; = &\;\ \bV^{(k)} + \bA^{(k+1)} - \bZ^{(k+1)}
	\end{align}
\end{subequations}
% \begin{subequations}
% 	\begin{align}
% 	\label{eq:pnp_stepa} \big\{\hat{\bM\;}, \hat{\bA\ } \big\} = &\mathop{\rm argmin}_{\bM, \{\balpha_i\}_{i=1}^N} \sum_{i=1}^N \mathcal{L}\big(\by_i,\hat{\by}_i\big) + \rho \|\bA-\hat{\bZ\, }\|^2 \qquad {\rm with } \quad   \hat{\by}_i = {\mathcal{F}}(\bM, \balpha_i) \\ \nonumber \qquad\;\;\;\; & {\rm s.t. } \quad   \bM\in\Omega_M, \, {\rm and \;} \balpha_i\in\Omega_\alpha\\ 
% 	\label{eq:pnp_stepb} {\hat{\bZ\,}}\; = &\;\ \mathop{\rm argmin}_{\bZ} \rho \|\bZ-{\hat{\bA\ }}\|^2 + \mathcal{R}(\bZ) \; = \; \texttt{Denoiser}\big({\hat{\bA\ }}\big)
% 	\end{align}
% \end{subequations}
% \begin{subequations}
% 	\begin{align}
% 	\label{eq:pnp_stepa} \big\{\hat{\bM}, \hat{\bA\; } \big\}_{k+1}= &\mathop{\rm argmin}_{\bM, \{\balpha_i\}_{i=1}^N} \sum_{i=1}^N \mathcal{L}\big(\by_i,\hat{\by}_i\big) + \rho \|\bA-\hat{\;\bZ_k}\|^2 \qquad {\rm with } \quad   \hat{\by}_i = {\mathcal{F}}(\bM, \balpha_i) \\ \nonumber \qquad\;\;\;\; & {\rm s.t. } \quad   \bM\in\Omega_M, \, {\rm and \;} \balpha_i\in\Omega_\alpha\\ 
% 	\label{eq:pnp_stepb} {\hat{\bZ}}_{k+1}\; = &\;\ \mathop{\rm argmin}_{\bZ} \rho \|\bZ-{\hat{\bA\ }}_{k+1}\|^2 + \mathcal{R}(\bZ) \; = \; \texttt{Denoiser}\big({\hat{\bA\ }}_{k+1}\big)
% 	\end{align}
% \end{subequations}
where superscript $^{(k)}$ refers to the iteration index, $\rho$ is the penalty parameter and $\bV^{(k)}$ is the dual variable to be updated at iteration $k$. In this formulation, the regularizer $\mathcal{R}\big(\bA\big)$ can be implicitly defined by a plugged denoising operator $\texttt{Denoiser}$, and~\eqref{eq:stepb_pnp} can be seen as the denoising {procedure} of $\big(\bA^{(k+1)}+\bV^{(k)}\big)$ and replaced by:
\boxedeq{eq_denoiser}{
{\bZ}^{(k+1)} =  \texttt{Denoiser}\big(\bA^{(k+1)}+\bV^{(k)}\big)
}

Generally, $\texttt{Denoiser}$ can be any off-the-shelf denoising operators. This offers the opportunity of incorporating CNN-based denoisers with powerful prior learning ability into the physical model-based iterative optimization. In particular, the work~\cite{zhao2021plug} considers the LMM assumption in~\eqref{eq_lmm} with known $\bM$ and incorporates conventional or deep learning based denoisers with two strategies, namely, a direct regularizer on $\bA$, and a regularizer on reconstructed spectra, i.e., $\bM\bA$. Considering the case of an unknown $\bM$, the work in~\cite{zhao2021hyperspectralGRSL} jointly estimates $\bM$ and exploits a handcrafted sparsity prior with the regularizer $||\bA||_{2,1}$, and a denoiser prior with implicit regularizer $\mathcal{R}\big(\bA\big)$ in the plug-and-play framework to produce more stable unmixing results. The 
work in~\cite{wang2020hyperspectral} proposes a nonlinear plug-and-play unmixing method based on the general bilinear model in~\eqref{eq_bilinear}, and both traditional and CNN-based denoisers are plugged into the iterative optimization.

\subsection{Unrolling iterative optimization into learnable deep architectures}
%Above denoisers in the plug-and-play framework, are closely black-box, which lacks physical interpretability and hinders the theoretical analysis of the unmixing mechanism. Regularized unmixing methods usually construct on an explicit model and have specific interpretable regularizers of latent variables. Deep learning based methods are more flexible function approximators and can directly learn priors from data. 
%Recent deep unrolling algorithms open a another window for integrating data-driven information in the regularized problem solving process. 

Different from plug-and-play algorithms, the deep unrolling paradigm unfolds and truncates iterative optimization algorithms into trainable deep architectures. Considering the LMM in~\eqref{eq_lmm} with known $\bM$, deep unrolling methods solve the unmixing problem in~\eqref{eq_FCLS} with the $||\bA||_{1}$ sparsity regularizer using iterative algorithms such as the iterative soft thresholding algorithm (ISTA) and ADMM {algorithm}. In~\cite{zhou2021admm}, by introducing an auxiliary variable $\bz$ such that $\bz = \balpha$ along with a dual variable $\bv$, the optimization problem is iteratively solved by the ADMM with the following steps:
\begin{subequations}
	\begin{align}
	\label{eq:stepa_unrolling} \balpha^{(k+1)} \; = &\; 
	%(\bM^\top\bM + \mu \bI)^{-1}(\bA^\top\by + \rho (\bz^{(k)} + \bv^{(k)})) \; = \; 
	\bW\by + \bB (\bz^{(k)} + \bv^{(k)}) \\
	\label{eq:stepb_unrolling} \bz^{(k+1)} \; = &\; \text{max}(\text{soft}\big(\balpha^{(k+1)} - \bv^{(k)}, {\lambda}/{\rho}\big), 0) 
	%\; = \; \text{max}(\textit{soft}\big(\ba^{(k+1)} - \bv^{(k)}, \theta\big), 0)
	\\
	\label{eq:stepc_unrolling} \bv^{(k+1)} \; = &\; \bv^{(k)} - \eta(\balpha^{(k+1)} - \bz^{(k+1)})
	\end{align}
\end{subequations}
% \begin{subequations}
% 	\begin{align}
% 	\label{eq:stepa} \hat{\bM\,} \; = &\; \mathop{\rm argmin}_{\bM} \|\bY-\bM\hat{\bA\ }\|_F^2 \quad {\rm s.t. } \quad   \bM\in\Omega_M \\ 
% 	\label{eq:stepb} \hat{\bA\ } \; = &\; \mathop{\rm argmin}_{\bA} \|\bY-\hat{\bM\,}\bA\|_F^2 + \lambda||\bA||_{p} \quad {\rm s.t. } \quad  \balpha_i\in\Omega_\alpha
% 	\end{align}
% \end{subequations}
where $\lambda$ is the regularization parameter, $\rho$ denotes the penalty parameter, $\eta$ is a parameter offering additional
flexibility, and $\bW = (\bM^\top\bM + \mu \bI)^{-1}\bM^\top$ and $\bB = (\bM^\top\bM + \mu \bI)^{-1}\rho$. Operator $\text{soft}(\cdot)$ in~\eqref{eq:stepb_unrolling} is the soft-threshold operator given by $\text{soft}(\bx, \lambda/\rho) = \text{sign}(\bx)(|\bx| - \lambda/\rho)_+$.  
% Based on multiplicative or proximal gradient descent methods, the updates of $\bM^{(k+1)}$ and $\bA^{(k+1)}$ boil down to iterative update steps with respect to $\{\bM^{(k)}, \bA^{(k)}\}$ and $\{\bM^{(k+1)}, \bA^{(k)}\}$. 
The key feature of deep unrolling unmixing algorithms is to leverage a reparameterization of these iterative update steps through the  consecutive {DNN} layers (denoted as {$f_{\ba}$, $f_{\bz}$ and $f_{\bv}$}) with learnable parameters $\{\bW^{(k+1)}, \bB^{(k+1)}, \theta^{(k+1)}, \eta^{(k+1)}\}$, respectively:
% \boxedeq{}{
% \begin{split}
% 	\bM^{(k+1)} \; = &\; \texttt{Layers}\{\bM^{(k)}, \bA^{(k)}; \Theta_{\bf M}^{(k)}\} \\ 
% 	\bA^{(k+1)} \; = &\; \texttt{Layers}\{\bM^{(k+1)}, \bA^{(k)}; \Theta_{\bf A}^{(k)}\}
% \end{split}}
\boxedeq{}{
\begin{split}
	\balpha^{(k+1)} \; = &\; f_{\ba}(\bz^{(k)}, \bv^{(k)}, \by; \bW^{(k+1)}, \bB^{(k+1)}) {\;=\; \bW^{(k+1)}\by + \bB^{(k+1)} (\bz^{(k)} + \bv^{(k)})}\\ 
	\bz^{(k+1)} \; = &\; f_{\bz}(\balpha^{(k+1)}, \bv^{(k)}; \theta^{(k+1)}) {\;=\; \text{ReLU}(\balpha^{(k+1)} - \bv^{(k)} -  \theta^{(k+1)}\bI)} \\
	\bv^{(k+1)} \; = &\; f_{\bv}(\balpha^{(k+1)}, \bz^{(k+1)}, \bv^{(k)}; \eta^{(k+1)}) {\;=\; \bv^{(k)} - \eta^{(k+1)}(\balpha^{(k+1)} - \bz^{(k+1)})}
\end{split}}
where $\theta^{(k+1)}$ is a learnable parameter which plays the role of $\lambda/\rho$ {and $\text{ReLU}(\cdot)$ is a component-wise rectified linear unit operation.}
These specific-designed layers are then cascaded into an entire DNN architecture for unmixing, which can be trained with supervised loss functions in an end-to-end manner. 
% Considering $||\bA||_{1}$, Qian et al.~\cite{qian2020spectral} solved the sub-problems in~\eqref{eq:stepa} and~\eqref{eq:stepb} by the multiplicative method and designed an DNN sharing an auto-encoder structure. More specifically, the encoder is constructed by unfolding the multiplicative iterations above, it maps the spectral pixels into a latent vector (abundances). The weights of the decoder are a learnable endmember matrix and restore the input spectra using the encoded data. 
For the blind unmixing problem in~\eqref{eq_FCLS_NMF}, the work in~\cite{xiong2021snmf} builds a non-convex sparse NMF model using $||\bA||_{p}$ with $0<p< 1$ and designs an interpretable sparsity constrained NMF network based on the ISTA to jointly estimate $\bM$ and $\balpha$. 

\section{Integrating loss learnt from data into physics-based inverse problems}
Loss functions $\mathcal{L}$ in \eqref{eq_Prob1}, \eqref{eq_Prob2} and \eqref{eq_Prob3} also play a crucial role in determining the unmixing performance. In addition to the data-driven model, data-learned loss functions have been proposed recently to improve the unmixing performance, particularly for Problem formulation-III.  The subsequent review focuses on autoencoder-based works as they mostly refer to Problem formulation-III.

\subsection{General geometric distances as cost functions}
Geometric distances which are usually used as the essential loss in autoencoder based unmixing methods enforce the network to reconstruct the input image. The works~\cite{su2019daen,ozkan2018end,qu2018udas} use the mean-squared error (MSE) to measure the similarity between the input and reconstructed pixels, that is,
\begin{equation}
\mathcal{L}_{\mathrm{MSE}}=\frac{1}{N} \sum_{i=1}^{N}\big\|\by_{i}-\hat{\by}_{i}\big\|^{2}. 
\end{equation}
In contrast to the MSE which is sensitive to the scale of spectra, the spectral angle distance (SAD) and spectral information divergence (SID) are scale-invariant. The SAD considers the angle between two spectra as a spectral similarity measure:
\begin{equation}
\mathcal{L}_{\mathrm{SAD}}=\frac{1}{N} \sum_{i=1}^{N} \arccos \left(\frac{\left\langle\by_{i}, \hat{\by}_{i}\right\rangle}{\left\|\by_{i}\right\|\left\|\hat{\by}_{i}\right\|}\right).
\end{equation}
The SID models the spectra as a probability distribution and considers the band-to-band spectral variability as the uncertainty existing in random variables to lower the gap between input and reconstructed pixels,
\begin{equation}
\mathcal{L}_{\mathrm{SID}}=\frac{1}{N} \sum_{i=1}^{N} \boldsymbol{p}_{i} \log \left(\frac{\boldsymbol{p}_{i}}{\hat{\boldsymbol{p}}_{i}}\right),
\end{equation}
where $\boldsymbol{p}=\left(\by / \mathbf{1}^{\top} \by\right)$ and $\hat{\boldsymbol{p}}=\left(\hat{\by} / \mathbf{1}^{\top} \hat{\by}\right)$ represent the probability distribution vector of the input and estimated pixels respectively. Nevertheless, these conventional losses are defined at the pixel level based on basic statistical assumptions that disregard inherent image structures.

\subsection{{Deep learning metrics as cost functions}}
Generative adversarial networks (GANs) have recently been used to address the limitations of geometric distance loss functions. A GAN consists of two components, a generator and a discriminator, and models the distribution of data using deep neural networks. The generator $\mathcal{G}$ is trained to generate samples that are similar to real data in order to confuse the discriminator. The discriminator $\mathcal{D}$ receives samples from the real data distribution (positive data) and samples from the generator (negative data) and attempts to distinguish between the real and generated samples. These two networks engage in a min-max adversarial game during the training process and they can be expressed as follows:
\begin{equation}
\min_{\mathcal{G}} \max_{\mathcal{D}} \mathrm{E}_{\bz \sim q(\bz|\by)}[\log \mathcal{D}(\bz)]+\mathrm{E}_{\bz\sim p(\bz)}[\log (1-\mathcal{D}(\mathcal{G}(\boldsymbol{z}))].
\end{equation}
%Using a discrimination network for unmixing can transfer the potentially intrinsic properties into consideration.
The architecture of unmixing methods using adversarial loss usually contains two terms namely, an autoencoder and a discriminator. The former, considered to be the generator, aims to estimate the abundances and extract the endmembers. The latter is a deep binary classifier and is used to distinguish between generated and real samples. Thus, the discriminator may be able to transfer intrusive properties of real data that are useful for unmixing. The loss function utilized for generator training is always composed of two terms:
\begin{equation}
    \mathcal{L}_{\mathcal{G}}=\mathcal{L}_{\mathrm{AE}}+\mathcal{L}_{\mathrm{adv}}
\end{equation}
{where $\mathcal{L}_{\mathrm{adv}}=\frac{1}{N} \sum_{i=1}^{N} \log \mathcal{D}\left(\bz_{i}\right)+\frac{1}{N} \sum_{i=1}^{N} \log \left(1-\mathcal{D}\left(\hat{\bz}_{i}\right)\right)$ is the loss function that is used to generate samples $\hat{\bz}$ that confuse the discriminator, and
$\mathcal{L}_{\mathrm{AE}}$ is the reconstruction loss function, e.g., the geometric cost function.}
The loss function for training the discriminator is defined as $\mathcal{L}_{\mathrm{D}}=-\mathcal{L}_{\mathrm{adv}}$.
The work~\cite{min2021jmnet} introduces the Wasserstein distance calculated from the discriminator as a regularization term to characterize the distribution similarity between the input and reconstructed spectra. %3 works
In~\cite{jin2021adversarial}, the adversarial network is utilized to model the abundance distribution.
%\noindent\textbf{Deep metric learning loss as cost function}

Recent works have introduced the perception mechanism to discover the discrepancy between reconstructed pixels and their corresponding ground-truth (input pixels). This strategy allows us to relax pixel-level reconstruction and enhance the unmixing performance using an end-to-end process. The perceptual similarity can be viewed as a feature matching operator that compares the perceptual features of the reconstructed output pixels to those of the input pixels. The expression for the perceptual loss defined in a feature domain is:
\begin{equation}
\mathcal{L}_{\mathrm{Perceptual}}=\frac{1}{N} \sum_{i=1}^{N}\left\|\mathcal{P}(\by_{i})-\mathcal{P}(\hat{\by}_{i})\right\|^{2},
\end{equation}
where $\mathcal{P}$ stands for a feature extractor.
For example, in~\cite{gao2021cycu}, a cycle-consistency strategy is used to further refine the detailed information in the unmixing process. In~\cite{min2021jmnet}, the features extracted from the hidden layers of a discriminator are used to ensure the consistency of high-level representations. 

\begin{sidewaystable}[]
\caption{{Features of the surveyed methods.}}\label{Tab_characteristics}
\vspace{3mm}
\renewcommand\arraystretch{0.7}
\centering
\scriptsize
\begin{tabular}{c|cccccc|l}
\hline\hline
\rowcolor[HTML]{C6E0B4} 
\textbf{} & \multicolumn{6}{c|}{\cellcolor[HTML]{C6E0B4}\textbf{Features}} & \multicolumn{1}{c}{\cellcolor[HTML]{C6E0B4}{\textbf{General pros and cons}}} \\ \hline
\rowcolor[HTML]{FFE699} 
\textbf{Reference} & \multicolumn{1}{c|}{\cellcolor[HTML]{FFE699}\textbf{Architecture}} & \multicolumn{1}{c|}{\cellcolor[HTML]{FFE699}\textbf{Model}} & \multicolumn{1}{c|}{\cellcolor[HTML]{FFE699}\textbf{Loss}} & \multicolumn{1}{c|}{\cellcolor[HTML]{FFE699}\textbf{Blind}} & \multicolumn{1}{c|}{\cellcolor[HTML]{FFE699}\textbf{Initialization}} & \textbf{Prior} & \cellcolor[HTML]{DDEBF7} \\ \cline{1-7}
\rowcolor[HTML]{DDEBF7} 
\cite{ozkan2018end} & \multicolumn{1}{c|}{\cellcolor[HTML]{DDEBF7}shallow layer} & \multicolumn{1}{c|}{\cellcolor[HTML]{DDEBF7}} & \multicolumn{1}{c|}{\cellcolor[HTML]{DDEBF7}MSE+SAD} & \multicolumn{1}{c|}{\cellcolor[HTML]{DDEBF7}} & \multicolumn{1}{c|}{\cellcolor[HTML]{DDEBF7}VCA} & \begin{tabular}[c]{@{}c@{}}$\ell_1$ on abundances,\\ $\ell_2$ on endmembers\end{tabular} & \cellcolor[HTML]{DDEBF7} \\ \cline{1-2} \cline{4-4} \cline{6-7}
\rowcolor[HTML]{DDEBF7} 
\cite{qu2018udas} & \multicolumn{1}{c|}{\cellcolor[HTML]{DDEBF7}\begin{tabular}[c]{@{}c@{}}untied denoising \\  autoencoder\end{tabular}} & \multicolumn{1}{c|}{\cellcolor[HTML]{DDEBF7}} & \multicolumn{1}{c|}{\cellcolor[HTML]{DDEBF7}MSE} & \multicolumn{1}{c|}{\cellcolor[HTML]{DDEBF7}} & \multicolumn{1}{c|}{\cellcolor[HTML]{DDEBF7}VCA+FCLS} & \begin{tabular}[c]{@{}c@{}}$\ell_{2,1}$ on\\  encoder weights\end{tabular} & \cellcolor[HTML]{DDEBF7} \\ \cline{1-2} \cline{4-4} \cline{6-7}
\rowcolor[HTML]{DDEBF7} 
\cite{su2019daen} & \multicolumn{1}{c|}{\cellcolor[HTML]{DDEBF7}\begin{tabular}[c]{@{}c@{}}stacked  autoencoder, \\ VAE\end{tabular}} & \multicolumn{1}{c|}{\cellcolor[HTML]{DDEBF7}} & \multicolumn{1}{c|}{\cellcolor[HTML]{DDEBF7}MSE} & \multicolumn{1}{c|}{\cellcolor[HTML]{DDEBF7}} & \multicolumn{1}{c|}{\cellcolor[HTML]{DDEBF7}VCA+FCLS} & minimum volume & \cellcolor[HTML]{DDEBF7} \\ \cline{1-2} \cline{4-4} \cline{6-7}
\rowcolor[HTML]{DDEBF7} 
\cite{palsson2018hyperspectral} & \multicolumn{1}{c|}{\cellcolor[HTML]{DDEBF7}deep encoder} & \multicolumn{1}{c|}{\cellcolor[HTML]{DDEBF7}} & \multicolumn{1}{c|}{\cellcolor[HTML]{DDEBF7}SAD} & \multicolumn{1}{c|}{\cellcolor[HTML]{DDEBF7}} & \multicolumn{1}{c|}{\cellcolor[HTML]{DDEBF7}VCA+FCLS} & n/a & \cellcolor[HTML]{DDEBF7} \\ \cline{1-2} \cline{4-4} \cline{6-7}
\rowcolor[HTML]{DDEBF7} 
\cite{palsson2020convolutional} & \multicolumn{1}{c|}{\cellcolor[HTML]{DDEBF7}CNN encoder} & \multicolumn{1}{c|}{\cellcolor[HTML]{DDEBF7}} & \multicolumn{1}{c|}{\cellcolor[HTML]{DDEBF7}SAD} & \multicolumn{1}{c|}{\multirow{-5}{*}{\cellcolor[HTML]{DDEBF7}$\surd$}} & \multicolumn{1}{c|}{\cellcolor[HTML]{DDEBF7}randomly} & \cellcolor[HTML]{DDEBF7}n/a & \cellcolor[HTML]{DDEBF7} \\ \cline{1-2} \cline{4-7}
\rowcolor[HTML]{DDEBF7} 
\cite{khajehrayeni2020hyperspectral} & \multicolumn{1}{c|}{\cellcolor[HTML]{DDEBF7}3D CNN encoder} & \multicolumn{1}{c|}{\multirow{-6}{*}{\cellcolor[HTML]{DDEBF7}LMM}} & \multicolumn{1}{c|}{\cellcolor[HTML]{DDEBF7}SID} & \multicolumn{1}{c|}{\cellcolor[HTML]{DDEBF7}$\times$} & \multicolumn{1}{c|}{\cellcolor[HTML]{DDEBF7}randomly} & \cellcolor[HTML]{DDEBF7}n/a & \cellcolor[HTML]{DDEBF7} \\ \cline{1-7}
\rowcolor[HTML]{DDEBF7} 
\cite{NAE2019} & \multicolumn{1}{c|}{\cellcolor[HTML]{DDEBF7}post-nonlinear decoder} & \multicolumn{1}{c|}{\cellcolor[HTML]{DDEBF7}} & \multicolumn{1}{c|}{\cellcolor[HTML]{DDEBF7}MSE} & \multicolumn{1}{c|}{\cellcolor[HTML]{DDEBF7}} & \multicolumn{1}{c|}{\cellcolor[HTML]{DDEBF7}VCA} & n/a & \cellcolor[HTML]{DDEBF7} \\ \cline{1-2} \cline{4-4} \cline{6-7}
\rowcolor[HTML]{DDEBF7} 
\cite{zhao2021lstm} & \multicolumn{1}{c|}{\cellcolor[HTML]{DDEBF7}\begin{tabular}[c]{@{}c@{}}LSTM encoder, additive \\ post-nonlinear decoder\end{tabular}} & \multicolumn{1}{c|}{\cellcolor[HTML]{DDEBF7}} & \multicolumn{1}{c|}{\cellcolor[HTML]{DDEBF7}MSE} & \multicolumn{1}{c|}{\cellcolor[HTML]{DDEBF7}} & \multicolumn{1}{c|}{\cellcolor[HTML]{DDEBF7}VCA} & TV on abundances & \cellcolor[HTML]{DDEBF7} \\ \cline{1-2} \cline{4-4} \cline{6-7}
\rowcolor[HTML]{DDEBF7} 
\cite{zhao2021hyperspectral} & \multicolumn{1}{c|}{\cellcolor[HTML]{DDEBF7}\begin{tabular}[c]{@{}c@{}}3D   CNN encoder, \\ generalized additive \\ nonlinear decoder\end{tabular}} & \multicolumn{1}{c|}{\cellcolor[HTML]{DDEBF7}} & \multicolumn{1}{c|}{\cellcolor[HTML]{DDEBF7}MSE} & \multicolumn{1}{c|}{\cellcolor[HTML]{DDEBF7}} & \multicolumn{1}{c|}{\cellcolor[HTML]{DDEBF7}VCA} & \begin{tabular}[c]{@{}c@{}}TV on endmembers,\\ $\ell_2$ weight decay\end{tabular} & \cellcolor[HTML]{DDEBF7} \\ \cline{1-2} \cline{4-4} \cline{6-7}
\rowcolor[HTML]{DDEBF7} 
\cite{su2020deep} & \multicolumn{1}{c|}{\cellcolor[HTML]{DDEBF7}\begin{tabular}[c]{@{}c@{}}multitask learning, \\ bilinear decoder\end{tabular}} & \multicolumn{1}{c|}{\cellcolor[HTML]{DDEBF7}} & \multicolumn{1}{c|}{\cellcolor[HTML]{DDEBF7}MSE} & \multicolumn{1}{c|}{\cellcolor[HTML]{DDEBF7}} & \multicolumn{1}{c|}{\cellcolor[HTML]{DDEBF7}VCA+FCLS} & {\color[HTML]{000000} \begin{tabular}[c]{@{}c@{}}graph regularizer\\ on abundances\end{tabular}} & \cellcolor[HTML]{DDEBF7} \\ \cline{1-2} \cline{4-4} \cline{6-7}
\rowcolor[HTML]{DDEBF7} 
\cite{shahid2021unsupervised} & \multicolumn{1}{c|}{\cellcolor[HTML]{DDEBF7}\begin{tabular}[c]{@{}c@{}}kernel-layer based encoder, \\ bilinear decoder,  \\ post-nonlinear decoder\end{tabular}} & \multicolumn{1}{c|}{\multirow{-5}{*}{\cellcolor[HTML]{DDEBF7}NLMM}} & \multicolumn{1}{c|}{\cellcolor[HTML]{DDEBF7}MSE} & \multicolumn{1}{c|}{\cellcolor[HTML]{DDEBF7}} & \multicolumn{1}{c|}{\cellcolor[HTML]{DDEBF7}K-means centers} & n/a & \cellcolor[HTML]{DDEBF7} \\ \cline{1-4} \cline{6-7}
\rowcolor[HTML]{DDEBF7} 
\cite{jin2021tanet} & \multicolumn{1}{c|}{\cellcolor[HTML]{DDEBF7}two-stream autoencoder} & \multicolumn{1}{c|}{\cellcolor[HTML]{DDEBF7}LMM} & \multicolumn{1}{c|}{\cellcolor[HTML]{DDEBF7}SID-SAD} & \multicolumn{1}{c|}{\cellcolor[HTML]{DDEBF7}} & \multicolumn{1}{c|}{\cellcolor[HTML]{DDEBF7}region-based VCA} & {\color[HTML]{000000} TV on pixel} & \cellcolor[HTML]{DDEBF7} \\ \cline{1-4} \cline{6-7}
\rowcolor[HTML]{DDEBF7} 
\cite{shi2021probabilistic} & \multicolumn{1}{c|}{\cellcolor[HTML]{DDEBF7}} & \multicolumn{1}{c|}{\cellcolor[HTML]{DDEBF7}} & \multicolumn{1}{c|}{\cellcolor[HTML]{DDEBF7}MSE+KL+SAD} & \multicolumn{1}{c|}{\cellcolor[HTML]{DDEBF7}} & \multicolumn{1}{c|}{\cellcolor[HTML]{DDEBF7}VCA} & minimum volume & \cellcolor[HTML]{DDEBF7} \\ \cline{1-1} \cline{4-4} \cline{6-7}
\rowcolor[HTML]{DDEBF7} 
\cite{borsoi2019deep} & \multicolumn{1}{c|}{\multirow{-2}{*}{\cellcolor[HTML]{DDEBF7}VAE}} & \multicolumn{1}{c|}{\multirow{-2}{*}{\cellcolor[HTML]{DDEBF7}\begin{tabular}[c]{@{}c@{}}spectral   \\ variability\end{tabular}}} & \multicolumn{1}{c|}{\cellcolor[HTML]{DDEBF7}MSE+KL} & \multicolumn{1}{c|}{\cellcolor[HTML]{DDEBF7}} & \multicolumn{1}{c|}{\cellcolor[HTML]{DDEBF7}FCLS} & TV on abundances & \cellcolor[HTML]{DDEBF7} \\ \cline{1-4} \cline{6-7}
\rowcolor[HTML]{DDEBF7} 
\cite{han2020deep} & \multicolumn{1}{c|}{\cellcolor[HTML]{DDEBF7}multi-branch encoder} & \multicolumn{1}{c|}{\cellcolor[HTML]{DDEBF7}NLMM} & \multicolumn{1}{c|}{\cellcolor[HTML]{DDEBF7}MSE} & \multicolumn{1}{c|}{\cellcolor[HTML]{DDEBF7}} & \multicolumn{1}{c|}{\cellcolor[HTML]{DDEBF7}\begin{tabular}[c]{@{}c@{}}scaled constrained\\  least squares (SCLS)\end{tabular}} & $\ell_2$ weight decay & \cellcolor[HTML]{DDEBF7} \\ \cline{1-4} \cline{6-7}
\rowcolor[HTML]{DDEBF7} 
\cite{hua2021dual} & \multicolumn{1}{c|}{\cellcolor[HTML]{DDEBF7}dual branch encoder} & \multicolumn{1}{c|}{\cellcolor[HTML]{DDEBF7}} & \multicolumn{1}{c|}{\cellcolor[HTML]{DDEBF7}\begin{tabular}[c]{@{}c@{}}squared sine \\ distance\end{tabular}} & \multicolumn{1}{c|}{\cellcolor[HTML]{DDEBF7}} & \multicolumn{1}{c|}{\cellcolor[HTML]{DDEBF7}VCA} & $\ell_{\frac{1}{2}}$ on abundances & \cellcolor[HTML]{DDEBF7} \\ \cline{1-2} \cline{4-4} \cline{6-7}
\rowcolor[HTML]{DDEBF7} 
\cite{gao2021cycu} & \multicolumn{1}{c|}{\cellcolor[HTML]{DDEBF7}cascaded autoencoder} & \multicolumn{1}{c|}{\cellcolor[HTML]{DDEBF7}} & \multicolumn{1}{c|}{\cellcolor[HTML]{DDEBF7}\begin{tabular}[c]{@{}c@{}}cycle-consistency \\ loss\end{tabular}} & \multicolumn{1}{c|}{\cellcolor[HTML]{DDEBF7}} & \multicolumn{1}{c|}{\cellcolor[HTML]{DDEBF7}VCA} & ASC constraint & \cellcolor[HTML]{DDEBF7} \\ \cline{1-2} \cline{4-4} \cline{6-7}
\rowcolor[HTML]{DDEBF7} 
\cite{rasti2021undip} & \multicolumn{1}{c|}{\cellcolor[HTML]{DDEBF7}deep image prior} & \multicolumn{1}{c|}{\cellcolor[HTML]{DDEBF7}} & \multicolumn{1}{c|}{\cellcolor[HTML]{DDEBF7}MSE} & \multicolumn{1}{c|}{\cellcolor[HTML]{DDEBF7}} & \multicolumn{1}{c|}{\cellcolor[HTML]{DDEBF7}\begin{tabular}[c]{@{}c@{}}a geometrical endmembers\\  estimation\end{tabular}} & n/a & \cellcolor[HTML]{DDEBF7} \\ \cline{1-2} \cline{4-4} \cline{6-7}
\rowcolor[HTML]{DDEBF7} 
\cite{qi2022sscu} & \multicolumn{1}{c|}{\cellcolor[HTML]{DDEBF7}\begin{tabular}[c]{@{}c@{}}two-stream autoencoder,\\ collaborative learning\end{tabular}} & \multicolumn{1}{c|}{\cellcolor[HTML]{DDEBF7}} & \multicolumn{1}{c|}{\cellcolor[HTML]{DDEBF7}SAD} & \multicolumn{1}{c|}{\cellcolor[HTML]{DDEBF7}} & \multicolumn{1}{c|}{\cellcolor[HTML]{DDEBF7}VCA+FCLS} & \cellcolor[HTML]{DDEBF7}$\ell_{\frac{1}{2}}$ on abundances & \cellcolor[HTML]{DDEBF7} \\ \cline{1-2} \cline{4-4} \cline{6-7}
\rowcolor[HTML]{DDEBF7} 
\cite{han2022multimodal} & \multicolumn{1}{c|}{\cellcolor[HTML]{DDEBF7}\begin{tabular}[c]{@{}c@{}}attention networks,\\ multimodal feature encoder\end{tabular}} & \multicolumn{1}{c|}{\cellcolor[HTML]{DDEBF7}} & \multicolumn{1}{c|}{\cellcolor[HTML]{DDEBF7}SAD} & \multicolumn{1}{c|}{\multirow{-14}{*}{\cellcolor[HTML]{DDEBF7}$\surd$}} & \multicolumn{1}{c|}{\cellcolor[HTML]{DDEBF7}VCA} & \cellcolor[HTML]{DDEBF7}$\ell_{\frac{1}{2}}$ on abundances & \cellcolor[HTML]{DDEBF7} \\ \cline{1-2} \cline{4-7}
\rowcolor[HTML]{DDEBF7} 
\cite{zhao2021plug} & \multicolumn{1}{c|}{\cellcolor[HTML]{DDEBF7}} & \multicolumn{1}{c|}{\cellcolor[HTML]{DDEBF7}} & \multicolumn{1}{c|}{\cellcolor[HTML]{DDEBF7}MSE} & \multicolumn{1}{c|}{\cellcolor[HTML]{DDEBF7}$\times$} & \multicolumn{1}{c|}{\cellcolor[HTML]{DDEBF7}FCLS} & denoiser prior & \cellcolor[HTML]{DDEBF7} \\ \cline{1-1} \cline{4-7}
\rowcolor[HTML]{DDEBF7} 
\cite{zhao2021hyperspectralGRSL} & \multicolumn{1}{c|}{\cellcolor[HTML]{DDEBF7}} & \multicolumn{1}{c|}{\multirow{-7}{*}{\cellcolor[HTML]{DDEBF7}LMM}} & \multicolumn{1}{c|}{\cellcolor[HTML]{DDEBF7}MSE} & \multicolumn{1}{c|}{\cellcolor[HTML]{DDEBF7}$\surd$} & \multicolumn{1}{c|}{\cellcolor[HTML]{DDEBF7}VCA+FCLS} & \cellcolor[HTML]{DDEBF7}\begin{tabular}[c]{@{}c@{}}denoiser prior,\\ $\ell_{2,1}$ on abundance\end{tabular} & \cellcolor[HTML]{DDEBF7} \\ \cline{1-1} \cline{3-7}
\rowcolor[HTML]{DDEBF7} 
\cite{wang2020hyperspectral} & \multicolumn{1}{c|}{\multirow{-3}{*}{\cellcolor[HTML]{DDEBF7}plug-and-play}} & \multicolumn{1}{c|}{\cellcolor[HTML]{DDEBF7}NLMM} & \multicolumn{1}{c|}{\cellcolor[HTML]{DDEBF7}MSE} & \multicolumn{1}{c|}{\cellcolor[HTML]{DDEBF7}$\times$} & \multicolumn{1}{c|}{\cellcolor[HTML]{DDEBF7}FCLS} & \cellcolor[HTML]{DDEBF7}denoiser prior & \cellcolor[HTML]{DDEBF7} \\ \cline{1-7}
\rowcolor[HTML]{DDEBF7} 
\cite{zhou2021admm} & \multicolumn{1}{c|}{\cellcolor[HTML]{DDEBF7}unroll ADMM} & \multicolumn{1}{c|}{\cellcolor[HTML]{DDEBF7}} & \multicolumn{1}{c|}{\cellcolor[HTML]{DDEBF7}MSE+SAD+SID} & \multicolumn{1}{c|}{\cellcolor[HTML]{DDEBF7}} & \multicolumn{1}{c|}{\cellcolor[HTML]{DDEBF7}VCA} & $\ell_1$ on abundances & \cellcolor[HTML]{DDEBF7} \\ \cline{1-2} \cline{4-4} \cline{6-7}
\rowcolor[HTML]{DDEBF7} 
\cite{xiong2021snmf} & \multicolumn{1}{c|}{\cellcolor[HTML]{DDEBF7}unroll NMF} & \multicolumn{1}{c|}{\cellcolor[HTML]{DDEBF7}} & \multicolumn{1}{c|}{\cellcolor[HTML]{DDEBF7}MSE} & \multicolumn{1}{c|}{\cellcolor[HTML]{DDEBF7}} & \multicolumn{1}{c|}{\cellcolor[HTML]{DDEBF7}VCA+FCLS} & $\ell_p$ on abundances & \cellcolor[HTML]{DDEBF7} \\ \cline{1-2} \cline{4-4} \cline{6-7}
\rowcolor[HTML]{DDEBF7} 
\cite{min2021jmnet} & \multicolumn{1}{c|}{\cellcolor[HTML]{DDEBF7}deep learning metric loss} & \multicolumn{1}{c|}{\cellcolor[HTML]{DDEBF7}} & \multicolumn{1}{c|}{\cellcolor[HTML]{DDEBF7}\begin{tabular}[c]{@{}c@{}}SAD+\\ Wasserstein  distance+\\ perceptual loss\end{tabular}} & \multicolumn{1}{c|}{\cellcolor[HTML]{DDEBF7}} & \multicolumn{1}{c|}{\cellcolor[HTML]{DDEBF7}randomly} & n/a & \cellcolor[HTML]{DDEBF7} \\ \cline{1-2} \cline{4-4} \cline{6-7}
\rowcolor[HTML]{DDEBF7} 
\cite{jin2021adversarial} & \multicolumn{1}{c|}{\cellcolor[HTML]{DDEBF7}adversarial network} & \multicolumn{1}{c|}{\multirow{-4}{*}{\cellcolor[HTML]{DDEBF7}LMM}} & \multicolumn{1}{c|}{\cellcolor[HTML]{DDEBF7}SAD+adversarial loss} & \multicolumn{1}{c|}{\multirow{-4}{*}{\cellcolor[HTML]{DDEBF7}$\surd$}} & \multicolumn{1}{c|}{\cellcolor[HTML]{DDEBF7}VCA+FCLS} & n/a & \multirow{-45}{*}{\cellcolor[HTML]{DDEBF7}\begin{tabular}[c]{@{}l@{}}\textbf{Pro 1.} Integrating  physics-based\\  models in decoder design enables\\  the methods to model complex\\  mixture mechanism with\\  physical characteristics.\\ \textbf{Pro 2.} Convolutional filters\\  are able to capture the spatial\\  correlation. \\ \textbf{Pro 3.} Integrating geometric\\  distance with deep learning\\  metric loss excavates\\  inherent image structures.\\ \textbf{Pro 4.} Proper initialization\\  can speed up the convergence\\  of the parameters and decrease\\  the possibility of parameter\\  overshooting.\\ \\ \textbf{Con 1.} Current methods require\\  to pre-selecting the model type\\  in advance and cannot involve\\  multiple mixing models in\\  complex scenes.\\ \textbf{Con 2.} For spatial discontinuity\\  regions, convolutional filters may\\  blur the abundance maps.\\ \textbf{Con 3.} Additional feature extraction\\  networks of calculating the deep\\  metric loss increase the number\\  of network parameters.\end{tabular}} \\ \hline\hline
\end{tabular}
\end{sidewaystable}

\section{Discussions and perspectives}

This article reviewed hyperspectral unmixing methods with special attention focused on the integration of physics-based and data-driven methods. A mathematical framework was established to introduce a variety of existing works with a unified formalism. As summarized in this paper, notable advances have been recently achieved in this topic.
{Table~\ref{Tab_characteristics} lists some features of the surveyed methods and summarizes some pros and cons of these methods.}
However, there are still several important aspects that merit further discussions and investigations.

\noindent\textbf{Performance evaluation:}
Evaluating the performance of hyperspectral unmixing algorithms has always been nontrivial since few research works have been devoted to producing ground-truth data~\cite{zhao2019laboratory}. Performance evaluation becomes even more difficult when considering data-driven methods. Preliminary comparison results of several autoencoder-based unmixing methods with synthetic and real datasets have been recently provided in~\cite{palsson2022blind}. This study demonstrates the significance of using appropriate loss functions, nonlinear methods, spatial information, and proper initialization. However, thorough and objective evaluation studies are still missing.

\noindent\textbf{Model selection:} Real data usually involve multiple mixing models in complex scenes. It is anticipated that data-driven models, such as the generalized linear-mixture/nonlinear fluctuation model, will exhibit superior scene-specific adaptability. Nevertheless, it is challenging, if not impossible, to pre-select an optimal model in advance. Using a mixing model with parallel multiple branches and an attention module to balance the contribution of each mixing model is one solution to this issue. Utilizing superpixel-based strategies and adopting distinct models within each superpixel and at its borders is an alternative solution.

\noindent\textbf{A fully integrated framework:} This article reviews the use of autoencoders to learn mixture models from data, the plug-and-play method to learn prior from data, and feature level losses from data. Existing works include one or two of above. Investigating an elegant strategy combining physics-based and data-driven methods that benefits from all of these factors is still missing. This requires deeper understanding and  original methods in both aspects of optimization and neural network design.

\noindent\textbf{Fast implementation:} Compared to classical optimization methods, data-driven approaches using {DNNs} require significantly higher computational resources especially when multiple data-driven modules collaborate to perform unmixing. Fast implementation must be considered for practical use of unmixing algorithms. The formulation of the optimization problems, the design of the network structures, and the way of integrating them are crucial for enhancing the convergence speed of the learning process. A trained encoder can be used for a supervised or semi-supervised unmixing when the computation resource is limited. Further, a compute unified device architecture (CUDA) implementation is also useful for executing parallel optimization-solving modules.

\bibliographystyle{IEEEbib}
\bibliography{BIB}

\begin{thebibliography}{10}

\bibitem{ghamisi2017advances}
P.~Ghamisi, N.~Yokoya, J.~Li, W.~Liao, S.~Liu, J.~Plaza, B.~Rasti, and
  A.~Plaza,
\newblock ``Advances in hyperspectral image and signal processing: A
  comprehensive overview of the state of the art,''
\newblock {\em IEEE Geosci. Remote Sens. Mag.}, vol. 5, no. 4, pp. 37--78,
  2017.

\bibitem{dobigeon2013nonlinear}
N.~Dobigeon, J.~Tourneret, C.~Richard, J.~Bermudez, S.~McLaughlin, and A.~Hero,
\newblock ``Nonlinear unmixing of hyperspectral images: Models and
  algorithms,''
\newblock {\em IEEE Signal Process. Mag.}, vol. 31, no. 1, pp. 82--94, 2013.

\bibitem{borsoi2021spectral}
R.~Borsoi, T.~Imbiriba, J.~Bermudez, C.~Richard, J.~Chanussot, L.~Drumetz,
  J.~Tourneret, A.~Zare, and C.~Jutten,
\newblock ``Spectral variability in hyperspectral data unmixing: A
  comprehensive review,''
\newblock {\em IEEE Geosci. Remote Sens. Mag.}, vol. 9, no. 4, pp. 223--270,
  2021.

\bibitem{chen2013nonlinear_sp}
J.~Chen, C.~Richard, and P.~Honeine,
\newblock ``Nonlinear estimation of material abundances in hyperspectral images
  with $\ell_1 $-norm spatial regularization,''
\newblock {\em IEEE Trans. Geosci. Remote Sens.}, vol. 52, no. 5, pp.
  2654--2665, 2013.

\bibitem{Heylen2014A}
R.~Heylen, M.~Parente, and P.~Gader,
\newblock ``A review of nonlinear hyperspectral unmixing methods,''
\newblock {\em IEEE J. Sel. Top. Appl. Earth Observat. Remote Sens.}, vol. 7,
  no. 6, pp. 1844--1868, 2014.

\bibitem{Halimi2011}
A.~Halimi, Y.~Altman, N.~Dobigeon, and J.-Y. Tourneret,
\newblock ``Nonlinear unmixing of hyperspectral images using a generalized
  bilinear model,''
\newblock {\em IEEE Trans. Geosci. Remote Sens.}, vol. 49, no. 11, pp.
  4153--4162, 2011.

\bibitem{hapke2012theory}
B.~Hapke,
\newblock {\em Theory of reflectance and emittance spectroscopy},
\newblock Cambridge university press, 2012.

\bibitem{heylen2011fully}
D.~Heinz and C.~Chang,
\newblock ``Fully constrained least squares linear spectral mixture analysis
  method for material quantification in hyperspectral imagery,''
\newblock {\em IEEE Trans. Geosci. Remote Sens.}, vol. 39, no. 3, pp. 529--545,
  2001.

\bibitem{miao2007endmember}
L.~Miao and H.~Qi,
\newblock ``Endmember extraction from highly mixed data using minimum volume
  constrained nonnegative matrix factorization,''
\newblock {\em IEEE Trans. Geosci. Remote Sens.}, vol. 45, no. 3, pp. 765--777,
  2007.

\bibitem{peng2021low}
J.~Peng, W.~Sun, H.~Li, W.~Li, X.~Meng, C.~Ge, and Q.~Du,
\newblock ``Low-rank and sparse representation for hyperspectral image
  processing: A review,''
\newblock {\em IEEE Geosci. Remote Sens. Mag.}, vol. 10, no. 1, pp. 10--43,
  2021.

\bibitem{li2019kernel}
Z.~Li, J.~Chen, and S.~Rahardja,
\newblock ``Kernel-based nonlinear spectral unmixing with dictionary pruning,''
\newblock {\em Remote Sensing}, vol. 11, no. 5, pp. 529, 2019.

\bibitem{heylen2010non}
R.~Heylen, D.~Burazerovic, and P.~Scheunders,
\newblock ``Non-linear spectral unmixing by geodesic simplex volume
  maximization,''
\newblock {\em IEEE J. Sel. Top. Sig. Process.}, vol. 5, no. 3, pp. 534--542,
  2010.

\bibitem{heylen2014distance}
R.~Heylen and P.~Scheunders,
\newblock ``A distance geometric framework for nonlinear hyperspectral
  unmixing,''
\newblock {\em IEEE J. Sel. Top. Appl. Earth Observat. Remote Sens.}, vol. 7,
  no. 6, pp. 1879--1888, 2014.

\bibitem{ozkan2018end}
S.~Ozkan, B.~Kaya, and G.~B. Akar,
\newblock ``{EndNet}: Sparse autoencoder network for endmember extraction and
  hyperspectral unmixing,''
\newblock {\em IEEE Trans. Geosci. Remote Sens.}, vol. 57, no. 1, pp. 482--496,
  2018.

\bibitem{qu2018udas}
Y.~Qu and H.~Qi,
\newblock ``{uDAS}: An untied denoising autoencoder with sparsity for spectral
  unmixing,''
\newblock {\em IEEE Trans. Geosci. Remote Sens.}, vol. 57, no. 3, pp.
  1698--1712, 2018.

\bibitem{su2019daen}
Y.~Su, J.~Li, A.~Plaza, A.~Marinoni, P.~Gamba, and S.~Chakravortty,
\newblock ``{DAEN}: Deep autoencoder networks for hyperspectral unmixing,''
\newblock {\em IEEE Trans. Geosci. Remote Sens.}, vol. 57, no. 7, pp.
  4309--4321, 2019.

\bibitem{palsson2018hyperspectral}
B.~Palsson, J.~Sigurdsson, J.~R. Sveinsson, and M.~O. Ulfarsson,
\newblock ``Hyperspectral unmixing using a neural network autoencoder,''
\newblock {\em IEEE Access}, vol. 6, pp. 25646--25656, 2018.

\bibitem{palsson2020convolutional}
B.~Palsson, M.~O. Ulfarsson, and J.~R. Sveinsson,
\newblock ``Convolutional autoencoder for spectral--spatial hyperspectral
  unmixing,''
\newblock {\em IEEE Trans. Geosci. Remote Sens.}, vol. 59, no. 1, pp. 535--549,
  2020.

\bibitem{khajehrayeni2020hyperspectral}
F.~Khajehrayeni and H.~Ghassemian,
\newblock ``Hyperspectral unmixing using deep convolutional autoencoders in a
  supervised scenario,''
\newblock {\em IEEE J. Sel. Top. Appl. Earth Observat. Remote Sens.}, vol. 13,
  pp. 567--576, 2020.

\bibitem{NAE2019}
M.~Wang, M.~Zhao, J.~Chen, and S.~Rahardja,
\newblock ``Nonlinear unmixing of hyperspectral data via deep autoencoder
  networks,''
\newblock {\em IEEE Geosci. Remote Sens. Lett.}, vol. 16, no. 9, pp.
  1467--1471, 2019.

\bibitem{zhao2021lstm}
M.~Zhao, L.~Yan, and J.~Chen,
\newblock ``{LSTM-DNN} based autoencoder network for nonlinear hyperspectral
  image unmixing,''
\newblock {\em IEEE J. Sel. Top. Sig. Process.}, vol. 15, no. 2, pp. 295--309,
  2021.

\bibitem{zhao2021hyperspectral}
M.~Zhao, M.~Wang, J.~Chen, and S.~Rahardja,
\newblock ``Hyperspectral unmixing for additive nonlinear models with a
  {3-D-CNN} autoencoder network,''
\newblock {\em IEEE Trans. Geosci. Remote Sens.}, vol. 60, pp. 1--15, 2022.

\bibitem{su2020deep}
Y.~Su, X.~Xu, J.~Li, H.~Qi, P.~Gamba, and A.~Plaza,
\newblock ``Deep autoencoders with multitask learning for bilinear
  hyperspectral unmixing,''
\newblock {\em IEEE Trans. Geosci. Remote Sens.}, vol. 59, no. 10, pp.
  8615--8629, 2020.

\bibitem{shahid2021unsupervised}
K.~T. Shahid and I.~D. Schizas,
\newblock ``Unsupervised hyperspectral unmixing via nonlinear autoencoders,''
\newblock {\em IEEE Trans. Geosci. Remote Sens.}, vol. 60, pp. 1--13, 2021.

\bibitem{jin2021tanet}
Q.~Jin, Y.~Ma, X.~Mei, and J.~Ma,
\newblock ``{TANet}: An unsupervised two-stream autoencoder network for
  hyperspectral unmixing,''
\newblock {\em IEEE Trans. Geosci. Remote Sens.}, vol. 60, pp. 1--15, 2021.

\bibitem{shi2021probabilistic}
S.~Shi, M.~Zhao, L.~Zhang, Y.~Altmann, and J.~Chen,
\newblock ``Probabilistic generative model for hyperspectral unmixing
  accounting for endmember variability,''
\newblock {\em IEEE Trans. Geosci. Remote Sens.}, vol. 60, pp. 1--15, 2022.

\bibitem{borsoi2019deep}
R.~Borsoi, T.~Imbiriba, and J.~Bermudez,
\newblock ``Deep generative endmember modeling: An application to unsupervised
  spectral unmixing,''
\newblock {\em IEEE Trans. Comput. Imag.}, vol. 6, pp. 374--384, 2019.

\bibitem{han2020deep}
Z.~Han, D.~Hong, L.~Gao, B.~Zhang, and J.~Chanussot,
\newblock ``Deep half-siamese networks for hyperspectral unmixing,''
\newblock {\em IEEE Geosci. Remote Sens. Lett.}, vol. 18, no. 11, pp.
  1996--2000, 2020.

\bibitem{hua2021dual}
Z.~Hua, X.~Li, Y.~Feng, and L.~Zhao,
\newblock ``Dual branch autoencoder network for spectral-spatial hyperspectral
  unmixing,''
\newblock {\em IEEE Geosci. Remote Sens. Lett.}, vol. 19, pp. 1--5, 2021.

\bibitem{gao2021cycu}
L.~Gao, Z.~Han, D.~Hong, B.~Zhang, and J.~Chanussot,
\newblock ``Cycu-net: Cycle-consistency unmixing network by learning cascaded
  autoencoders,''
\newblock {\em IEEE Trans. Geosci. Remote Sens.}, vol. 60, pp. 1--14, 2021.

\bibitem{rasti2021undip}
B.~Rasti, B.~Koirala, P.~Scheunders, and P.~Ghamisi,
\newblock ``{UnDIP}: Hyperspectral unmixing using deep image prior,''
\newblock {\em IEEE Trans. Geosci. Remote Sens.}, vol. 60, pp. 1--15, 2021.

\bibitem{qi2022sscu}
L.~Qi, F.~Gao, J.~Dong, X.~Gao, and Q.~Du,
\newblock ``{SSCU-Net}: Spatial--spectral collaborative unmixing network for
  hyperspectral images,''
\newblock {\em IEEE Trans. Geosci. Remote Sens.}, vol. 60, pp. 1--15, 2022.

\bibitem{han2022multimodal}
Z.~Han, D.~Hong, L.~Gao, J.~Yao, B.~Zhang, and J.~Chanussot,
\newblock ``Multimodal hyperspectral unmixing: Insights from attention
  networks,''
\newblock {\em IEEE Trans. Geosci. Remote Sens.}, vol. 60, pp. 1--13, 2022.

\bibitem{zhao2021plug}
M.~Zhao, X.~Wang, J.~Chen, and W.~Chen,
\newblock ``A plug-and-play priors framework for hyperspectral unmixing,''
\newblock {\em IEEE Trans. Geosci. Remote Sens.}, vol. 60, pp. 1--13, 2021.

\bibitem{zhao2021hyperspectralGRSL}
M.~Zhao, T.~Gao, J.~Chen, and W.~Chen,
\newblock ``Hyperspectral unmixing via nonnegative matrix factorization with
  handcrafted and learned priors,''
\newblock {\em IEEE Geosci. Remote Sens. Lett.}, vol. 19, pp. 1--5, 2021.

\bibitem{wang2020hyperspectral}
Z.~Wang, L.~Zhuang, L.~Gao, A.~Marinoni, B.~Zhang, and M.~K. Ng,
\newblock ``Hyperspectral nonlinear unmixing by using plug-and-play prior for
  abundance maps,''
\newblock {\em Remote Sensing}, vol. 12, no. 24, pp. 4117, 2020.

\bibitem{zhou2021admm}
C.~Zhou and M.~R. Rodrigues,
\newblock ``{ADMM}-based hyperspectral unmixing networks for abundance and
  endmember estimation,''
\newblock {\em IEEE Trans. Geosci. Remote Sens.}, vol. 60, pp. 1--18, 2021.

\bibitem{xiong2021snmf}
F.~Xiong, J.~Zhou, S.~Tao, J.~Lu, and Y.~Qian,
\newblock ``{SNMF-Net}: Learning a deep alternating neural network for
  hyperspectral unmixing,''
\newblock {\em IEEE Trans. Geosci. Remote Sens.}, vol. 60, pp. 1--16, 2021.

\bibitem{min2021jmnet}
A.~Min, Z.~Guo, H.~Li, and J.~Peng,
\newblock ``{JMnet}: Joint metric neural network for hyperspectral unmixing,''
\newblock {\em IEEE Trans. Geosci. Remote Sens.}, vol. 60, pp. 1--12, 2021.

\bibitem{jin2021adversarial}
Q.~Jin, Y.~Ma, F.~Fan, J.~Huang, X.~Mei, and J.~Ma,
\newblock ``Adversarial autoencoder network for hyperspectral unmixing,''
\newblock {\em IEEE Trans. Neu. Net. Learn. Sys.}, pp. 1--15, 2021.

\bibitem{zhao2019laboratory}
M.~Zhao, J.~Chen, and Z.~He,
\newblock ``A laboratory-created dataset with ground truth for hyperspectral
  unmixing evaluation,''
\newblock {\em IEEE J. Sel. Top. Appl. Earth Observat. Remote Sens.}, vol. 12,
  no. 7, pp. 2170--2183, 2019.

\bibitem{palsson2022blind}
B.~Palsson, J.~Sveinsson, and M.~Ulfarsson,
\newblock ``Blind hyperspectral unmixing using autoencoders: A critical
  comparison,''
\newblock {\em IEEE J. Sel. Top. Appl. Earth Observat. Remote Sens.}, vol. 15,
  pp. 1340--1372, 2022.

\end{thebibliography}

\end{document}